\documentclass[a4paper]{article}
\pdfoutput=1

\usepackage[utf8]{inputenc}
\usepackage[T1]{fontenc}
\usepackage[english]{babel}

\usepackage[pdftex,breaklinks=true,bookmarks=true,pdfborder={0 0 0},colorlinks=false]{hyperref}

\usepackage{listings}
\lstdefinelanguage{sl}{alsolanguage=C,morekeywords={sl_def,sl_enddef,sl_index,sl_parm,sl_getp,sl_setp,__asm__,sl_glparm,sl__static,sl_shparm,__volatile__,__typeof__,sl_create,sl_createsync,sl_glarg,sl_sharg,sl_geta,sl_sync,sl__exclusive,sl__static,sl_detach,sl_decl,sl_typedef_fptr,sl_glfparm,sl_glfarg,sl_shfparm,sl_shfarg,inline,sl_seta}}

\usepackage{cleveref}
\usepackage{graphicx}
\usepackage[caption=false]{subfig}

\newcommand{\ie}{i.e.~}
\newcommand{\eg}{e.g.~}
\newcommand{\cf}{cf.~}
\newcommand{\vs}{vs.~}

\usepackage{enumitem}
\setlist{nolistsep}
\usepackage{array}

\begin{document}

\author{M. Lankamp, R. Poss, Q. Yang, J. Fu, I. Uddin, C.R. Jesshope\\University of Amsterdam, The Netherlands}
\title{MGSim---simulation tools for multi-core processor architectures}

\maketitle

\begin{abstract}
MGSim is an open source discrete event simulator for on-chip hardware components,
developed at the University
of Amsterdam. It is intended to be a research and teaching vehicle to
study the fine-grained hardware/software interactions on many-core and
hardware multithreaded processors. It includes support for core models
with different instruction sets, a configurable
multi-core interconnect, multiple configurable cache and memory
models, a dedicated I/O subsystem, and comprehensive monitoring and
interaction facilities. The default model configuration shipped with
MGSim implements Microgrids, a many-core architecture
with hardware concurrency management.
MGSim is furthermore written mostly in C++ and uses
object classes to represent
chip components. It is optimized for 
architecture models that can be described as process networks.
\end{abstract}

\setcounter{tocdepth}{1}
\tableofcontents

\clearpage
\section{Introduction}

MGSim is a discrete event simulator for on-chip hardware components,
developed by a group of researchers at the CSA group\footnote{\url{http://csa.science.uva.nl/}} at the University
of Amsterdam since 2007.
MGSim is developed as a research and teaching vehicle to
study the fine-grained hardware/software interactions on many-core and
hardware multithreaded processors. It includes support for core models
with different Instruction Set Architectures (ISAs), a configurable
multi-core interconnect, multiple configurable cache and memory
models, a dedicated I/O subsystem, and comprehensive monitoring and
interaction facilities. The default model configuration shipped with
MGSim simulates the \emph{Microgrid platform}, so that programmers can
use MGSim as a full-system emulation of a computer equipped with a
Microgrid~\cite{poss.12.dsd}.

As a software infrastructure, MGSim's component models and simulation
kernel are written in C++; they use object classes to represent the
chip components. Ancillary tools
are written in Python. A characteristic
feature of the MGSim framework is that it promotes the definition of
architecture models where components across clock domains only
synchronize via FIFO buffers, \ie where models can be described as
process networks. MGSim is further available\footnote{currently hosted at \url{http://svp-dev.github.com/}.} free of charge
under an open source license.

This article reviews the MGSim tool box, as of version 3.3. We
start in \cref{sec:ctx} by describing its background story and how it
compares to its competitors. We then review its simulation framework
in \cref{sec:sim}, followed by its existing component models in
\cref{sec:comps}.  \Cref{sec:mon} then shows how the user can inspect
the state of the simulation. Finally, \cref{ssec:fut} outlines
possible future developments and \cref{sec:conc} wraps up the
presentation.

\section{Context and related work}\label{sec:ctx}

\subsection{Motivation and audience}

Historically, MGSim was first developed to explore the behavior of
D-RISC cores~\cite{bolychevsky.96.ieee,bernard.08.samos} when grouped
together in a many-core processor chip. Since then, MGSim has matured
into a versatile framework to simulate many-core architectures. The
development of MGSim is guided by two main purposes.

The first is \textbf{support scientific research in the design of many-core
  general-purpose processor architectures}.  So far, the
  implementation of MGSim has been influenced by the research
  activities of the CSA group at the University of Amsterdam, namely:
\begin{itemize}
\item design space exploration of multi-processor systems-on-chip, \ie testing
different combinations of architecture parameters to optimize platforms towards
specific applications, and
\item design of new techniques in processor micro-architecture, \ie adding or changing features
in individual processor cores and their memory interconnect to obtain higher performance/cost ratios.
\end{itemize}

The other purpose of MGSim is to \textbf{support undergraduate and
  graduate education activities in computer architecture, parallel
  programming, compiler construction and operating system design}. In particular,
MGSim was tailored to the following requirements:
\begin{itemize}
\item provide a human-scale software infrastructure that can be comprehended by standalone students in these fields,
\item integrate the emulation platform and its operating software in a software package that
 can be seamlessly deployed and get ready to run on student computers with minimal effort,
\item provide interfaces to observe and illustrate the internal workings of a system while it is running.
\end{itemize}

\subsection{Emulation \vs simulation}

MGSim can be considered both as a simulation framework and a
full-system emulator. 

Simulators are tools that reproduce abstract models of component
behaviors, often to make predictions about them. Simulators
are key items in the toolbox of hardware architects, who
need to explore the behavior of computing components
before their design is finalized, using 
pre-selected software benchmarks.

Emulators are tools that reproduce the concrete behavior
of a computing artifact, including the hardware/software
interface of a given processor. Emulators are used, for
example, during compiler and operating system development, to
debug software in a controlled environment before it
ships to production.

In short, simulators exist to debug and optimize hardware
designs, whereas emulators exist to debug and optimize software running
on the emulated system.

\begin{figure}
\centering
\includegraphics[width=.6\linewidth]{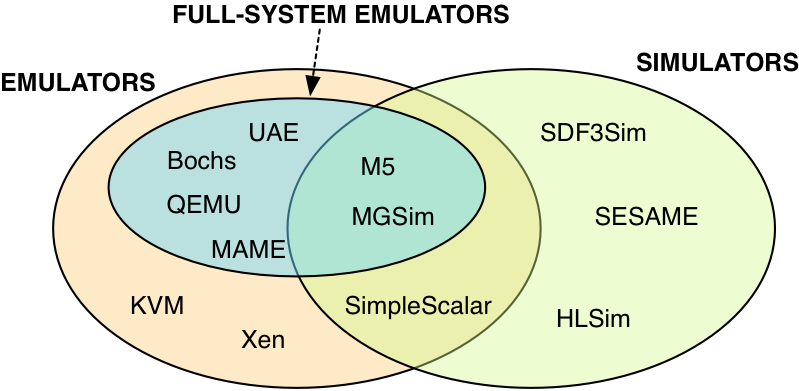}
\caption{Emulators and simulators: a Venn diagram with examples.}\label{fig:emusimu}
\end{figure}

Emulators can be further divided between partial and full system
emulation (also called ``virtualization''). With partial emulation,
only application code runs within the emulation environment, and
operating system functions are serviced through a host/guest
interface. With a full emulation, the entire software stack runs on
the emulated hardware. MGSim can serve both as a hardware simulator and full-system emulator.
We illustrate
these distinctions in \cref{fig:emusimu}.

\subsection{Related work}\label{sec:related}

In the group of software frameworks that are both simulators and
emulators, as illustrated above, MGSim most relates to
SimpleScalar\footnote{\url{http://simplescalar.com/}} and
Gem5\footnote{\url{http://m5sim.org}}~\cite{binkert.06.micro}.

In contrast to MGSim, SimpleScalar only provide partial emulation:
operating system functions are served on the host platform via the
\texttt{syscall} pseudo-instruction. MGSim was designed
as a full-system emulation so as to also study the behavior
of operating software when running over the emulated platform.
Moreover, SimpleScalar was primarily designed to emulate single-core
platforms, where MGSim's focus lies towards multi-core platforms.

Its purpose and even its software architecture make MGSim much closer
to the Gem5 framework. Gem5, like MGSim, consists of a library of C++
components that can be grouped in configurable topologies to define
multi-core platforms. Both frameworks are discrete event,
component-based simulations able to emulate full systems. At the time
of this writing, Gem5 even offers more monitoring and visualization
facilities than MGSim.

The differences between Gem5 and MGSim can be found at two levels.

Firstly, Gem5 was designed and motivated to emulate \emph{existing}
platforms. In particular, one of its design requirements was to be able
to run entire existing software stacks unchanged, for example
GNU/Linux, FreeBSD, L4K or Solaris. MGSim does not share this
requirement, and its implementation is thus much simpler than Gem5's.
This makes MGSim more accessible for education activities than Gem5.

Secondly, Gem5 started as a single-core system emulator, focusing on
the accurate simulation of large, state-of-the-art sequential
processors. Multi-core support was only added later, and Gem5 is still
optimized for use with few cores sharing a high-level functional
emulation of a cache coherency network and inter-processor interrupt
network. In contrast, MGSim was designed from the ground-up as a
many-core network featuring different detailed memory interconnects
and a dedicated point-to-point messaging network between cores.  This
makes MGSim a potentially more productive tool for research in
operating software for fine-grained multi-core applications.

\section{Architecture of the simulator}\label{sec:sim}

\begin{figure}
\centering
\includegraphics[width=.6\linewidth]{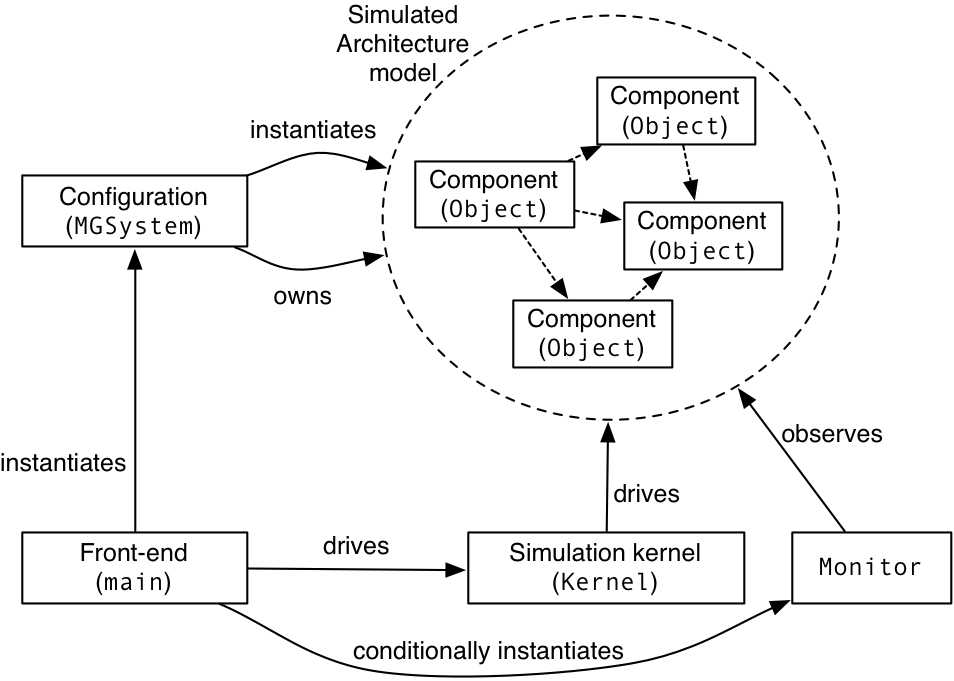}
\caption{Entity-relationship diagram of an MGSim simulator.}\label{fig:parts}
\end{figure}

MGSim's framework is composed of five main parts, illustrated in \cref{fig:parts}.

The \emph{simulation kernel} provides base abstractions for processes,
registers, buffers, latches, arbitrators and ports, as well as an
execution driver that schedules ready component processes at each
simulation cycle.

The \emph{library of component models} provides object classes for
the various component types found on the chip: processors (cores), 
caches, memory networks, I/O interconnects, etc. The component
models are implemented using the base abstractions from
the simulation kernel. Typically, a component will define one or more
processes, optionally some internal state for its processes, and
latches, buffers and/or arbitrators visible from other components.

The \emph{system configuration constructor} instantiates
the component models and connects them together to form a full
architecture model. This part is further distributed between a top-level
``system topology'' constructor and the individual constructors of
component models, which may choose to instantiate sub-components or
dependent components.

The \emph{simulation front-end} provides a user interface to
MGSim. The interface is composed of command-line parsing,
configuration file loader, interactive command interpreter, event
trace filtering, asynchronous monitoring, etc.

Finally, the optional \emph{asynchronous monitor} runs asynchronously
in a separate thread of execution. It periodically samples the state
of selected components and writes it to a trace file or FIFO for
analysis or visualization by external tools. See \cref{sec:mon} for details.

\subsection{Simulation overview}

\Cref{tab:start} provides a step-by-step overview of the
initialization of MGSim and the handling of one simulation step.

\begin{table}
\centering
\scriptsize
\begin{tabular}{l>{\ttfamily\raggedright\arraybackslash}l>{\ttfamily\raggedright\arraybackslash}l>{\ttfamily\raggedright\arraybackslash}l>{\ttfamily\raggedright\arraybackslash}l}
  & \normalfont Front-end & \normalfont Configuration & \normalfont Components & \normalfont Kernel \\
\hline
1. & main() & & &  \\
2. & $\leftrightarrow$Config::Config() & & & \\
3. & & $\rightarrow$MGSystem::MGSystem() & $\leftrightarrow$$\forall C. C$::$C$() & \\
4. & $\rightarrow$HandleCommandLine() & & & \\
5. & $\hookrightarrow$cmd\_run() & & & \\
6. & $\hookrightarrow$StepSystem() & & & $\rightarrow$Kernel::Step() \\
7. & & & & $\leftrightarrow$  $\forall S. S$::Update() \\
8. & & & $\forall P. P$::$\mathit{CycleHandler}$(Aq)$\leftrightarrow$ \\
9. & & & & $\leftrightarrow$  $\forall A. A$::OnArbitrate() \\
10. & & & $\forall P. P$::$\mathit{CycleHandler}$(Ck)$\leftrightarrow$ \\
11. & & & $\forall P. P$::$\mathit{CycleHandler}$(Ct)$\leftrightarrow$ \\
\end{tabular}
\caption{Call graph of MGSim's initialization and first simulation cycle.}\label{tab:start}
\end{table}

Upon initialization of an MGSim instance, the front-end parses the
command-line parameters and configuration file(s) (step 1). It then creates a
\texttt{Config} object that holds a database of configuration
variables (step 2). The front-end then instantiates the
configuration constructor (\texttt{MGSystem}) which in turn populates
the architecture model by instantiating components according to the
configuration (step 3). After this point the model is ready,
no further objects are constructed, and the simulation can start.

If invoked to run interactively, the front-end displays an interactive
prompt and accepts user commands (step 4). For example, invoking the
\texttt{run} command starts the simulation by triggering the
\texttt{Step} method of the simulation kernel (steps 5 \& 6). \texttt{Step}
can advance the simulation by one or more cycles. At every cycle, the following happens:
\begin{enumerate}[start=7]
\item Any pending updates to stateful structures shared by components (\eg FIFO buffers)
   are committed, to become visible during the new cycle. 
\item The \emph{acquire} phase of the cycle is run for all active component processes. 
   During the acquire phase, process handlers declare their intent to use shared
   structures and \emph{request arbitration}. During this phase, processes may not
   update internal state.
\item After the acquire phase completes, all involved arbitration requests
   are resolved by the kernel. 
\item Once arbitration has been resolved, all active processes
   run the \emph{check} phase of the cycle. During this phase,
   the \emph{results of arbitration is reported to each process}, which
   determines \emph{which control path to use} (\eg stall, access another shared storage, etc.). Again,
   during this phase, processes may not update internal state. 
\item Once the check phase has completed, all remaining non-blocked
   processes run the \emph{commit} phase of the cycle. During
   this phase, processes use the control path chosen during the check
   phase, may update their internal state, and declare updates 
   to shared storage to be effected at the start of the next cycle.
   They may also emit informational messages to be logged to
   a synchronous event trace by the kernel.
\end{enumerate}

At the end of each cycle, active processes are then rescheduled to run
at the next cycle or some cycles later, according to their simulated
clock frequency.

Note that during the check phase, processes may become blocked because of denied
arbitration, but also when attempting to read from empty FIFOs. When a
process becomes idle on an empty input FIFO, it will thus only be
reactivated after a subsequent cycle produces data into the FIFO. This
is the mechanism by which MGSim models the behavior of asynchronous
networks of components.

\subsection{Anatomy of a component}

Components in the simulation framework are intended to correspond
roughly to components on chip, \ie to an area of hardware. They are
organized in a tree, where each child node represents a sub-parts of
its parent component. For example, the \texttt{DCache} (L1 data cache)
and \texttt{Pipeline} components are child nodes of
the processor component (\texttt{Processor}) which encompasses them.

The base interface for all components is \texttt{Object}. Via \texttt{Object}, each
component is related to:
\begin{itemize}
\item a name that identifies it relative to its siblings in the tree;
\item its parent component and children components, if any;
\item a clock domain (either its own or shared with its parent);
\item the simulation kernel that drives the entire component tree.
\end{itemize}

Additionally, each component may define one or more of the following:
\begin{itemize}
\item \emph{processes}, which represent state machines or functional circuits;
\item \emph{shared storages and arbitrators}, \eg FIFO buffers,
  registers or single-bit latches, which may be used by two or more processes
  including processes from other components, and which may
  cause processes to block upon access;
\item \emph{internal state} used by only one process, or state shared
  by processes of the same component that does not require arbitration nor
  decides process scheduling;
\item \emph{services}, which provide part of the logic of processes
   from other components;
\item \emph{inspection handlers}, which are invoked from MGSim's
  interactive command prompt upon user commands,
\item \emph{administrative data} for the simulator itself which does
  not represent hardware components (\eg counters for statistics).
\end{itemize}   

We can illustrate these concepts using the \texttt{ICache} component
which models an L1 instruction cache. This defines, for example:

\begin{itemize}
\item the two processes \texttt{p\_Outgoing} and \texttt{p\_Incoming} for the 
  input and output queuing from and to memory, respectively;
\item the two buffers \texttt{m\_outgoing} and \texttt{m\_incoming} shared
  with the memory subsystem, which may control the scheduling of \texttt{p\_Outgoing}/\texttt{p\_Incoming};
\item the internal state \texttt{m\_lines} which represents the cache's data blocks,
   protected against concurrent accesses by the shared arbitrator \texttt{p\_service};
\item the service \texttt{Fetch}, used by the pipeline process of the same name and
  which writes to \texttt{m\_outgoing}, 
  and the service \texttt{OnMemoryReadCompleted}, used by the connected memory process
  upon completion of loads and which writes to \texttt{m\_incoming};
\item an administrative reference \texttt{m\_memory} to the simulation component in
  charge of simulating the memory system;
\item the administrative counter \texttt{m\_numHits} used to generate cache statistics.
\end{itemize}

\subsection{Component processes}\label{ssec:proc}

Processes in the simulation framework represent the activities of
data transformation and communication in the system. 

All processes are triggered by the availability of data in a specific
shared storage, which we will call its \emph{source storage}. When
triggered, a process becomes active and its \emph{cycle handler} is
called by the kernel at every cycle of the corresponding clock
domain. The process' cycle handler may then in turn attempt to acquire
more storage or arbitrators, fail while doing so and thus
\emph{stall}. When stalled, the cycle handler will re-try the same
behavior in subsequent cycles until the behavior succeeds. Upon
successful completion, a process may either \emph{consume data from its
source storage} or stay ready to be invoked again for another behavior
in the next cycle (\eg in state machines). A process becomes
\emph{idle} when its source storage becomes \emph{empty}.

The base interface for all processes is \texttt{Process}. Via \texttt{Process}, 
each process is related to:
\begin{itemize}
\item its enclosing component;
\item a name that identifies it within the enclosing component;
\item its state (running/stalled or idle);
\item its cycle handler invoked at every cycle of the clock domain while running or stalled.
\end{itemize}

\begin{lstlisting}[language=c++,float,caption={Behavior of the \texttt{ICache}'s \texttt{Outgoing} process.},label=lst:iout]
Result ICache::DoOutgoing()
{
    auto& address = m_outgoing.Front();
    if (!m_memory.Read(address))
    {
        DeadlockWrite("Unable to read");
        return FAILED;
    }
    m_outgoing.Pop();
    return SUCCESS;
}
\end{lstlisting}

We can illustrate these concepts using the \texttt{Outgoing} process
of the \texttt{ICache} component. This process is triggered by the
\texttt{m\_outgoing} FIFO buffer, which is populated with refill requests
by the pipeline's fetch stage upon I-cache misses. The
\texttt{Outgoing} process is in charge of picking the front-most
request at each cycle and sending it to the memory system. Its
(simplified) handler is given in \cref{lst:iout}.  The behavior is to
try to issue a memory load via the service
\texttt{m\_memory.Read()}. If the service is unsuccessful (\eg due to
memory contention), the handler in turn reports failure to the kernel,
which will cause the kernel to stall the process.  When stalled,
\texttt{Outgoing} handler will be called again at its next cycle. If
the request can be issued to memory, the handler then consumes the
entry with \texttt{m\_outgoing.Pop()} and reports success. If
\texttt{m\_outgoing} becomes empty as a result, the \texttt{Outgoing}
is then marked idle by the kernel.

The macro \texttt{DeadlockWrite} registers a message that is only
printed by the kernel if all processes in the system become stalled,
which would be a sign of deadlock.

Process instantiation occurs as part of the enclosing component's
instantiation. During instantiation, the component's constructor will
gives a name to the process and declares the process and its cycle
handler to the simulation kernel. The cycle handler is typically
implemented as a C++ method in the enclosing component, for example
\texttt{ICache::DoOutgoing()} in this example.

\begin{lstlisting}[language=c++,float,caption={Simplified behavior of \texttt{ICache}'s \texttt{Incoming} process.},label=lst:iin]
Result ICache::DoIncoming()
{
    if (!p_service.Invoke())
        return FAILED;

    auto   lineid  = m_incoming.Front();
    COMMIT{ m_lines[lineid].state = LINE_FULL; }
    // ...
    m_incoming.Pop();
    return SUCCESS;
}
\end{lstlisting}

Two additional features can be found in the handler of \texttt{ICache}'s other
process \texttt{Incoming}, given in \cref{lst:iin}. This
process is triggered by \texttt{m\_incoming}, a FIFO buffer
populated by memory load completion messages
coming from memory, \ie refill data after cache misses. 

Because this handler accesses the internal cache state
\texttt{m\_lines}, it must first acquire the arbitrator
\texttt{p\_service} which protects the line data against access races
with the pipeline's fetch stage (the \texttt{Fetch} service also
acquires \texttt{p\_service}).

Arbitrator acquisition occurs in two phases. During the
\emph{acquire} phase of the cycle, the handler
sees the \texttt{Invoke} method always succeed. However,
during that phase \texttt{Invoke} also registers the
arbitration request. After the acquire phase completes,
the kernel resolves arbitrations and marks processes
for which arbitration fails. Then
the process handler is run again during the \emph{check}
phase of the cycle, and \texttt{Invoke} returns
true or false depending on the arbitration results. 
 
Note that the entirety of the handler executes during the
\emph{acquire} phase ``as if'' arbitration had succeeded
already. However, internal component state must not be modified at
this phase, since the process may be arbitrated away for this cycle as
a result.  Therefore, any update to internal state, like
\texttt{m\_lines} in the example above, must be enclosed in a
\texttt{COMMIT} block which is only performed during the phase of the
same name in the cycle.

Accesses to shared storages via \texttt{Push} and \texttt{Pop} is
already internally handled via a \texttt{COMMIT} block and thus do not
need the extra marking in process handlers.

The execution of process handlers in three phases and the use of
\texttt{COMMIT} blocks for state updates is the fundamental mechanism
through which MGSim achieves transactions in components.

\subsection{Shared storage}

Shared storage is a basic abstractions of the simulation
kernel that represent state visible to one or more
processes, with controlled update semantics:

\begin{itemize}
\item processes updating the storage must be in the same clock domain;
\item if linked as a source storage to a process, the storage will
   activate the process when non-empty;
\item multiple updates in the same cycle are rejected (a form of arbitration)
   unless explicitly allowed by the storage model.
\end{itemize}   

The following sub-types of storage are provided:
\begin{itemize}
\item \texttt{Buffer}: a FIFO buffer with accessors \texttt{Push},
  \texttt{Pop}, \texttt{Front};
\item \texttt{Register}: a full/empty storage cell with accessors
  \texttt{Read}, \texttt{Write}, \texttt{Clear}, \texttt{Empty};
\item \texttt{SingleFlag}: a single bit trigger with accessors
  \texttt{Set}, \texttt{Clear}, \texttt{IsSet}.
\end{itemize}

Of these, \texttt{Buffer} is the only model that supports multiple
updates in the same cycle, up to a maximum set during
instantiation. This reflects the ability of most FIFO circuits to
process FIFO appends faster than the cycle time of their client
processes.

\subsection{Arbitrated services and ports}

When two or more processes must synchronize their access
to a shared storage or some component's internal state,
arbitration is required. For this MGSim provides two abstractions.

The first abstraction, \emph{arbitrated services}, model
standalone arbitration circuits, independent from the storage or state
that they protect. They are loosely analogous to the concept of
``mutex lock'' in concurrent programming languages; when they grant
access to a process, the process's handler can perform multiple
updates to the protected state atomically.  For example, the
arbitrator \texttt{p\_service} in \texttt{ICache} discussed previously
is an arbitrated service, protecting the internal state
\texttt{m\_lines} from concurrent access by the \texttt{Fetch} service
and \texttt{Incoming} process.

The second abstraction, \emph{arbitrated ports}, model
arbitration circuits combined with a storage structure with multiple
entries, for example a register file. Arbitrated ports arbitrate on
each entry of the structure separately, depending on the index
requested. This way, two accesses in the same cycle to different
indices do not cause their respective processes to stall; whereas two accesses to the same
index causes one of the processes to stall due to arbitration.

As of this writing, arbitrated ports are only used for the register
files of the core model \texttt{Processor}; other shared storage and
state is modeled as a single entity and thus only require the
protection offered by arbitrated services, which is cheaper to
simulate.

Furthermore, both arbitrated services and ports are parameterized with an
\emph{arbitration strategy}, which models which arbitration circuit to
use. The following arbitrators are available:
\begin{itemize}
\item \emph{priority} arbitrators: multiple processes
  are connected at fixed priorities; during a cycle, the
  process with the highest priority is granted access. This is the simplest and cheapest arbitrator in hardware, 
  but the priorities must be chosen carefully to avoid starvation and priority inversion.
\item \emph{cyclic} arbitrators: multiple processes are connected at
  equal priority; during a cycle, a counter is incremented and grants
  access in a round-robin fashion. This is slightly more expensive to
  implement in hardware, however it guarantees fairness.
\item \emph{priority-cyclic} arbitrators: akin to a priority arbitrator,
  with the lowest priority level shared by multiple processes using a cyclic arbitrator. This
  is the most expensive arbitrator to implement in hardware.
\end{itemize}
  
\subsection{Termination and deadlock detection}

Since processes are only activated when shared storage is updated,
and shared storage is updated only by other processes, 
a situation where all processes in the system become idle
corresponds to a system where there is nothing more to do, \ie
a halted machine. This situation is recognized by MGSim's
kernel and terminates the simulation without error.

Meanwhile, 
since MGSim is intended for use in architecture research,
a situation can occur where the architecture model contains
errors which cause the system to deadlock. In particular,
two or more processes may be erroneously defined to mutually wait on each
other by competing for access to a cyclical chain of shared storages.
As per \cref{ssec:proc} above, active processes waiting on arbitration
declare themselves to MGSim's kernel as \emph{stalled}. Therefore,
if the kernel detects during a cycle that all non-idle processes
have become stalled, it can conclude that a deadlock has occurred
and terminate the simulation with a deadlock error. 

Moreover, each individual cycle handler that marks its process as
stalling may register a ``deadlock message'' for the kernel with
\texttt{DeadlockWrite}. Upon detecting a deadlock, the kernel replays
all involved cycle handlers to actually print the deadlock messages,
so that the user can investigate the situation.

Intuitively, these mechanisms are limited in two situations:

\begin{itemize}
\item when a livelock occurs, the livelocked processes appear to
  MGSim's kernel as if they were making progress (they are non-idle
  and non-stalled), and the situation cannot be detected. It is
  expected that this form of error is handled at the software level
  running on the simulated platform by means of timeouts.
\item when the simulated platform is waiting on external input, the
  corresponding process in the simulated input device marks itself as
  running (\ie not idle and not stalled) until external input is
  received. If a deadlock occurs in the rest of the system during that
  time, MGSim's kernel is then unable to detect it. For this reason,
  all simulated I/O devices provide the concept of ``activation,''
  whereby the input process is only running when the device is
  activated.  Activation/deactivation is then requested explicitly by
  system software running on the simulated platform during input
  operations, to minimize the amount of time where deadlock detection is
  disabled.
\end{itemize}

\subsection{Contracts with the implementer}\label{sec:pitfalls}

The abstractions provided by MGSim's kernel model circuits templates
that are known to be constructable in hardware. However, the
implementer of components for MGSim may fall victim to three pitfalls
which are not protected against by the MGSim framework.

The first pitfall was mentioned above; it is the ability to define a
deadlocking architecture. This can happen when a group of two or more
processes are directly or indirectly waiting for each other via a
cyclical arbitration chain. To avoid this pitfall requires analysis of
the process dependency graph and thorough testing of modeled designs.

The second pitfall is the ability to express a cycle handler for a
process that is not implementable in hardware. As of this writing,
MGSim does not restrict which part of the C++ language can be used in
a process handler. It is thus possible to erroneously use the full
generality of the language, and define a behavior that has potentially
unbounded space and time requirements. To avoid this pitfall requires
checking via code analysis that the expressed computations correspond
to primitive recursive functions with existing equivalent circuit
implementations.

The third pitfall is the ability to configure a simulated clock
frequency incompatible with the length of the critical path through
the circuit implementing a process cycle handler. Indeed, any cycle
handler definition corresponds to some minimum complexity in
hardware. This complexity in turn places a lower bound on the duration
of a clock cycle, at a given silicon technology and energy
supply. MGSim does not prevent the implementer or user from defining a
simulated clock frequency with a smaller cycle duration than required
by the hardware circuit. To avoid this pitfall requires expert
knowledge about the relationship between the circuit specification in
a cycle handler, the circuit complexity in hardware, and the
relationship between energy, gate density, critical path length and
frequency for the envisioned silicon technology. 

This last pitfall is shared by most simulation frameworks that
simulate at a level higher than individual circuit gates. The first
two pitfalls may be partially prevented in the future by automated solutions, \cf
\cref{ssec:fut} for details.

\section{Component model library}\label{sec:comps}

MGSim is shipped with a set of predefined component models and a
predefined, configurable system topology. The preset library further
contains three component categories: \emph{processors}, \emph{I/O
  devices and interconnect}, and \emph{memory systems}.

\subsection{Processor/core models}

The MGSim library provides two predefined processor models:
\texttt{Processor} and \texttt{FPU}.  They are separate since one FPU
can be shared by two or more processors.

\begin{figure}
\centering
\includegraphics[width=.5\linewidth]{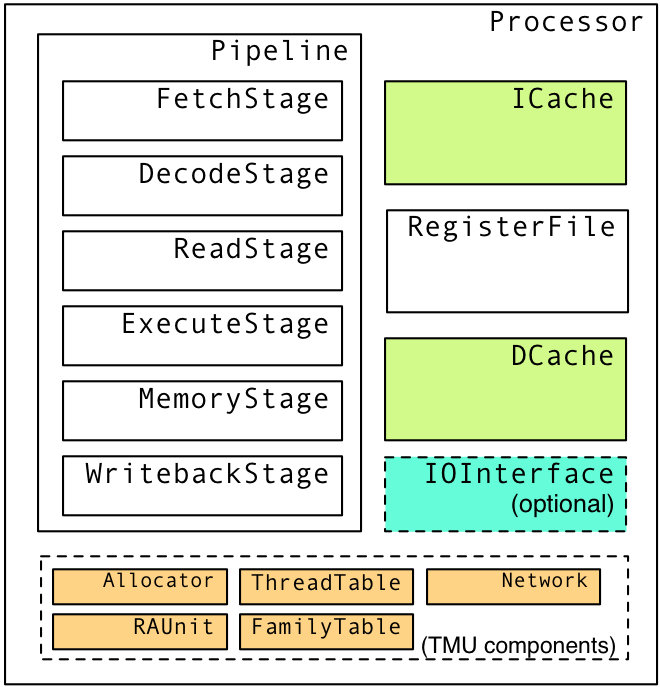}
\caption{Overview of \texttt{Processor} and its sub-components.}\label{fig:procarch}
\end{figure}

The \texttt{Processor} model implements a simple processor core with
an in-order, single-issue, six-stage RISC pipeline. An overview of its
sub-components is given in \cref{fig:procarch}. As of this writing, the
model can emulate a subset of either the DEC Alpha ISA (64-bit, little
endian), the SPARCv8 ISA (32-bit, big endian) or the MIPS I ISA
(32-bit, either little or big endian). All ISAs are emulated on top of
the same six pipeline stages, albeit with different data paths and
stage behavior. For example, the SPARC model consumes two cycles in
the read stage for a ``store double'' (\texttt{std}) instruction to
read its three register operands over the two read ports to the
register file, and consumes also two cycles in the memory stage for
double loads and stores.

In addition to the base pipeline behavior, the \texttt{Processor}
model also provides out-of-order completion for some instruction types
using dynamic dataflow scheduling. This is done by equipping the
register file with full/empty bits on each register, and marking the
output operand of long-latency instructions to be ``pending''. When a
subsequent instruction reads from a ``pending'' input, it is scheduled
to wait until the operand becomes available, \ie until the original
instruction completes. This feature enables latency tolerance, as \eg
loads to missing L1 cache lines do not stall the pipeline, and
independent subsequent instructions can continue to execute. This is a
feature provided by the D-RISC model that \texttt{Processor} was
originally designed to simulate. Details about the dynamic scheduling
abilities of D-RISC are detailed
in~\cite{lankamp.07,poss.12.dsd,lankamp.12}
and~\cite[Chap.~3]{poss.12}.

Furthermore, the \texttt{Processor} model can also simulate hardware
multithreading, up to a configurable number of hardware threads. This
feature is also inherited from the D-RISC model that
\texttt{Processor} was designed to simulate. The hardware scheduling
is similar to the one found in the Niagara~\cite{kongetira.05.micro}
architecture, although MGSim's \texttt{Processor} can also
preemptively switch threads at the fetch stage if an instruction is
known in advance to possibly miss its input dependency. Like Niagara,
it provides 0-cycle switch overhead.

\begin{figure}
\centering
\includegraphics[width=.8\linewidth]{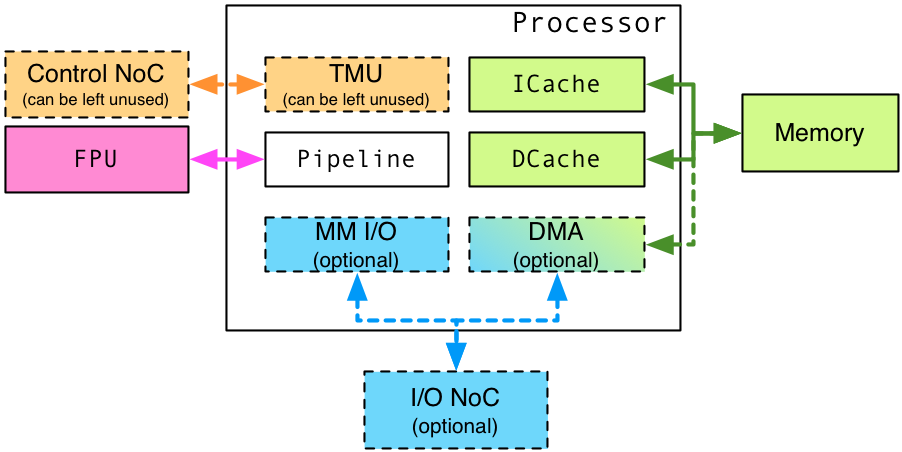}
\caption{External interfaces of the predefined \texttt{Processor} model.}\label{fig:procifs}
\end{figure}

The \texttt{Processor} model supports the following external interfaces, illustrated
in \cref{fig:procifs}:

\begin{itemize}
\item an interface to an instance of \texttt{FPU}, which may be
  shared with other processors. The interface is bidirectional
  and asynchronous: FP operations are dispatched in-order by \texttt{Processor}
  to the appropriate queue/pipeline on the FPU, but may complete out of order.
\item an interface to a memory system via its L1 data and instruction
  caches. If supported by the memory system, \texttt{Processor} can
  issue multiple requests before waiting for an answer, and even
  communicate bidirectionally.
\item an interface to a control network-on-chip (NoC), which may be
  separate from the memory system. Through this network,
  \texttt{Processor} can receive messages that influence its hardware
  threads or remotely access its registers via a dedicated Thread
  Management Unit (TMU) in hardware. From its D-RISC heritage,
  \texttt{Processor} also provides ISA extensions to send messages
  over this NoC, described in~\cite[Chap.4 \& App.~D]{poss.12}.
\item optionally, an interface to an I/O NoC, which may be separate
  from the two previous networks. When enabled, \texttt{Processor}
  simulates a memory-mapped interface at the memory stage of the
  pipeline to send and receive messages directly to/from I/O devices
  via this NoC. It also simulates an integrated Direct Memory Access
  (DMA) controller able to convert I/O messages from the NoC to memory
  requests without participation from the main pipeline.
\item administrative interfaces for inspection and debugging by the user of the
  MGSim framework.
\end{itemize}

At the time of this writing, the choice of ISA to be emulated by
\texttt{Processor} is static: a compiled MGSim instance can either
support an Alpha ISA, or a SPARC ISA, etc., but not both in the same
simulator executable. This choice was made to enable compile-time
optimizations around important architectural constants, namely: the
width of a data word, the width of memory addresses and data
endianness. Most other architectural parameters are configurable upon
MGSim's initialization at run-time, including L1 cache
sizes/associativity/indexing, the core's clock frequency, FPU-core
mappings, the availability of an I/O interface and the buffer sizes of
the memory/NoC interfaces.

\subsection{Memory models}

The MGSim library provides multiple memory systems to connect
processor cores to external memory. The existence of several memory
models side-by-side is motivated by ongoing scientific research in
multi-core architectures: for evaluation, practitioners typically
find it useful to perform comparative analyses of the behavior of
multi-core systems using different memory architectures and system
topologies.

Although MGSim's framework does not limit the applicability of each
model, not all combinations of memory models and number of cores are
accurate and useful. The complexity of the hardware circuits
corresponding to some models grows unfavorably with the number of
cores, suggesting to switch to another memory architecture instead.
Moreover, as outlined in \cref{sec:pitfalls}, the user should
care to align the frequency of the memory clock domain according
to the expected hardware complexity of the chosen configuration. 
At the time of this writing, MGSim's framework does not contain
logic to automatically scale frequency to circuit complexity nor
report unrealistic/unimplementable configurations; expert
knowledge by the user is expected to apply.

\subsubsection{Serial memory}\label{sec:sermem}

\begin{figure}[t]
\centering
\includegraphics[scale=.55]{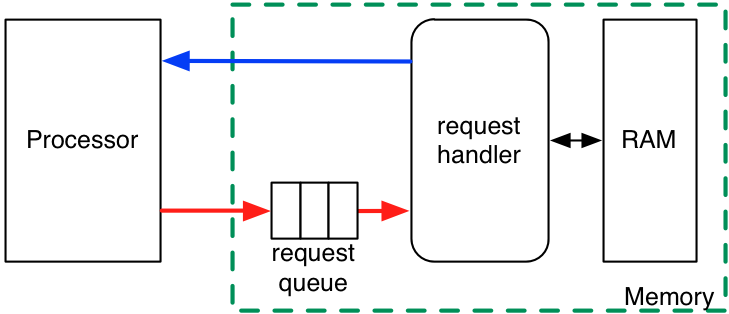}
\caption{Serial memory model (\texttt{SerialMemory}) with one processor.}\label{fig:sermem1}
\end{figure}

\begin{figure}[t]
\centering
\includegraphics[scale=.55]{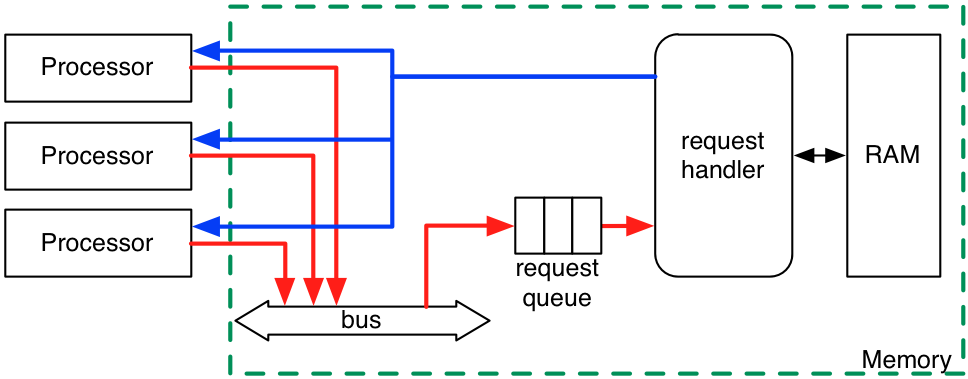}
\caption{Serial memory model in a multi-core configuration.}\label{fig:sermem}
\end{figure}

The simplest memory model is \texttt{SerialMemory}, illustrated in
\cref{fig:sermem1,fig:sermem}. 

In a single-core configuration, load and store requests from the
processor are queued in a buffer (16 entries by default) and handled
one after another by a request handler process, which simulates a
memory controller. The latency to handle each request is
configurable. The responses are served directly to the processor by
the request handler.

In a multi-core configuration, the components do not change but an
MGSim arbitrated service controls access to the request buffer,
thereby simulating a request bus between all processors. Responses to
requests are served directly to the processor that initiated the
request using a separate bus without arbitration (as there is only a single
initiator). Write requests snoop on the L1 cache of
all other processors on the bus (\ie after being granted access by the
arbitrator).

The memory systems that this model simulates are highly sensitive to
contention on the request bus, and thus rarely used practice with more
than 4-8 cores.

The front-end configuration key \texttt{MemoryType} can be set to
\texttt{SERIAL} to automatically build a memory system of this type.

\subsubsection{Parallel memory}

\begin{figure}[t]
\centering
\includegraphics[scale=.55]{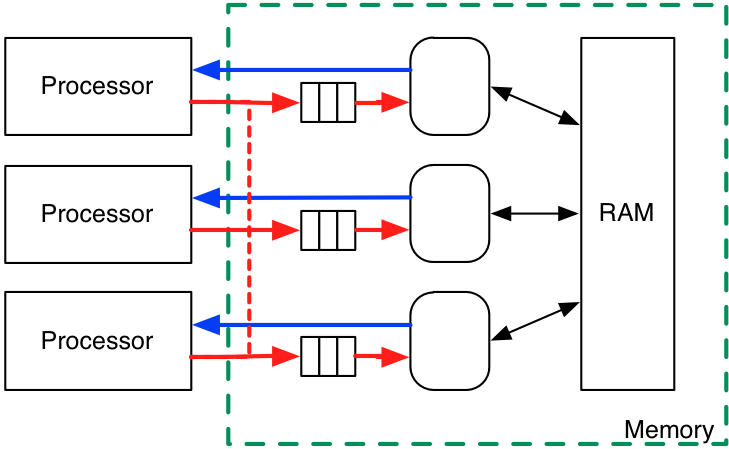}
\caption{Parallel memory model (\texttt{ParallelMemory}) in a multi-core configuration.}\label{fig:parmem}
\end{figure}

MGSim's \texttt{ParallelMemory} model is identical to
\texttt{SerialMemory} in a single-core configuration, but differs in a
multi-core configuration as illustrated in \cref{fig:parmem}. Here,
each processor has its own port to the memory, implemented as its own
request queue and handler. Thanks to this independence,
there is no contention between separate cores. Write
requests snoop on the L1 cache of all other processors transparently
(\ie snoop contention is not simulated).

This model is intended to represent an ideal parallel memory. It can
be used \eg to study the theoretical bounds on throughput and latency
when contention is not considered. It is not intended to be an
accurate model of hardware; a hardware implementation would require either an
arbitrator between the request handlers to access the physical memory
array, or to bank the memory and use one arbitrated port per bank.

The front-end configuration key \texttt{MemoryType} can be set to
\texttt{PARALLEL} to automatically build a memory system of this type.

\subsubsection{Banked memory in a single-core configuration}\label{sec:bankedmem}

\begin{figure}[t]
\centering 
\includegraphics[scale=.45]{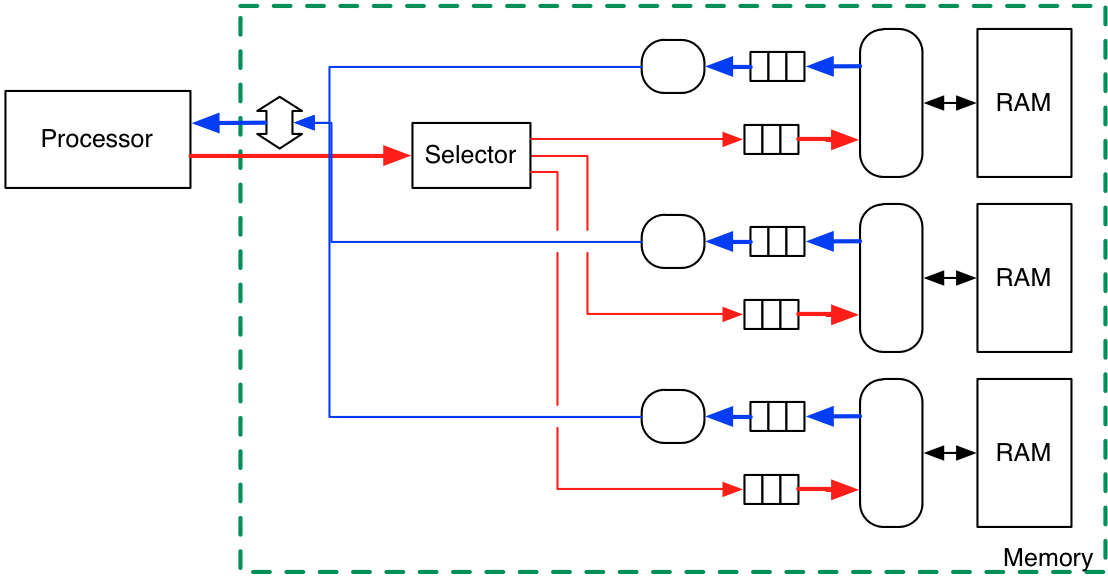}
\caption{Banked memory model (\texttt{BankedMemory}).}\label{fig:bankedmem1}
\end{figure}

\begin{figure}[t]
\centering
\includegraphics[scale=.45]{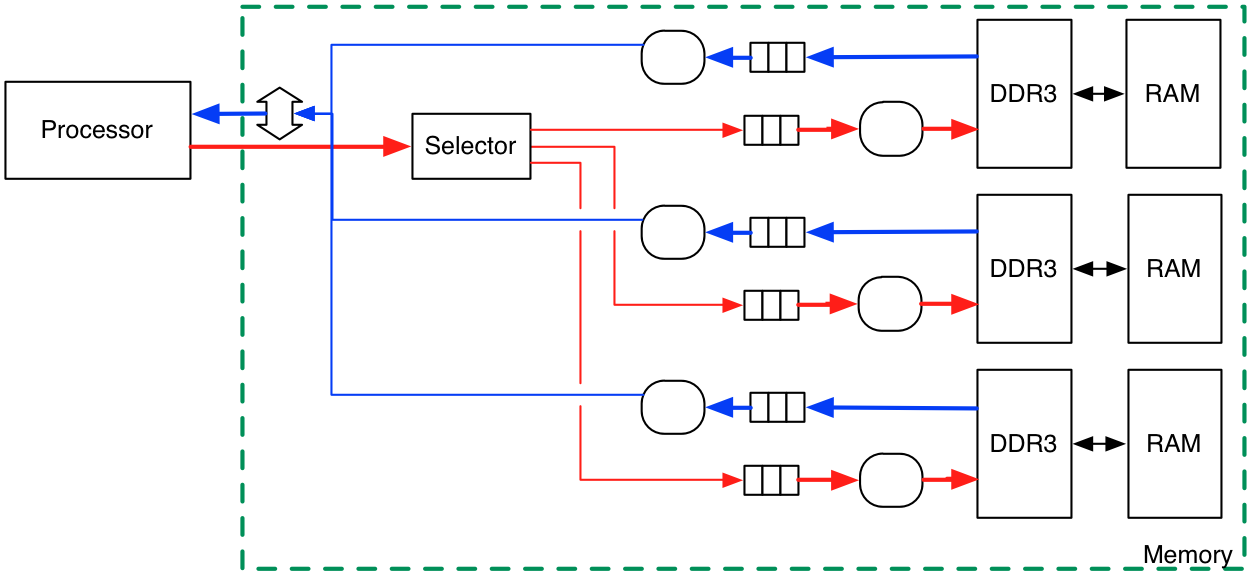}
\caption{Banked memory model with DDR3 controllers (\texttt{DDRMemory}).}\label{fig:ddrmem1}
\end{figure}

Two predefined models are provided that offer memory-level parallelism
by separating requests to different memory banks:
\texttt{BankedMemory} (\cref{fig:bankedmem1}) and \texttt{DDRMemory}
(\cref{fig:ddrmem1}).  The former uses a sequential request handler
like \texttt{SerialMemory} above, whereas the latter uses a simulation
of a DDR3 channel controller. The DDR3 protocol provides additional
parallelism by overlapping the handling of subsequent requests in a
pipeline.

In both models, the number of banks is configurable. The selection of
which bank or DDR channel to use is made for each request coming from
the processor based on the memory address. The request is then queued
at the selected bank/channel, processed asynchronously, and the
response is then queued for delivery. Responses from different
banks/channels to the same processor are delivered using an arbitrated
service, \ie at most once per memory clock cycle.

These models are intended to simulate hardware memory systems where
one processor is connected to multiple memory modules via a
bidirectional bus.

\subsubsection{Address to bank mappings}\label{sec:bankselectors}

Which \emph{selection strategy} is used to select banks/channels is
also configurable. MGSim's component library provides the following:

\begin{itemize}
\item \texttt{ZeroSelector}: always select the first bank/channel (used for troubleshooting);
\item \texttt{DirectSelector}, \texttt{DirectSelectorBinary}: uses the
  lowest order bits of the address as bank index;
\item \texttt{RightXOR}, \texttt{RightAdd}: computes the bank/channel index using
  an exclusive binary OR (for \texttt{RightXOR}) or sum (for
  \texttt{RightAdd}) of two low order blocks of index-sized bits in
  the address, as per~\cite{gonzalez.96.isc};
\item \texttt{XORFold}, \texttt{AddFold}: generalization of
  \texttt{RightXOR}/\texttt{RightAdd} to all bits of the address;
\item \texttt{RotationMix4}: compute a four-step, rotation-based hash
  of the address and use the low-order bits of the hash to select a
  bank/channel.
\end{itemize}

\texttt{DirectSelectorBinary} models the most common circuits in
hardware, but is sensitive to strided memory accesses that align with
the number of banks: these would cause contention on a single
bank. The model \texttt{RotationMix4} provides near-ideal
randomization of the address to bank mapping and can be thus
considered to simulate near-ideal load distribution. Its higher
circuit complexity makes it an unlikely hardware implementation,
though.

Four predefined settings are recognized by the front-end for 
configuration key \texttt{MemoryType}:
\begin{itemize}
\item \texttt{BANKED}: \texttt{BankedMemory} with either the
  \texttt{DirectSelector} or \texttt{DirectSelectorBinary} strategy,
  depending on the number of banks configured (the latter is used
  when the number of banks is a power of two);
\item \texttt{RANDOMBANKED}: \texttt{BankedMemory} with the
  \texttt{RotationMix4} strategy;
\item \texttt{DDR}: \texttt{DDRmemory} with either the
  \texttt{DirectSelector} or \texttt{DirectSelectorBinary} strategy,
  depending on the number of DDR channels configured;
\item \texttt{RANDOMDDR}: \texttt{DDRMemory} with the
  \texttt{RotationMix4} strategy.
\end{itemize}

\subsubsection{Banked memory in a multi-core configuration}

\begin{figure}[t]
\centering
\includegraphics[scale=.45]{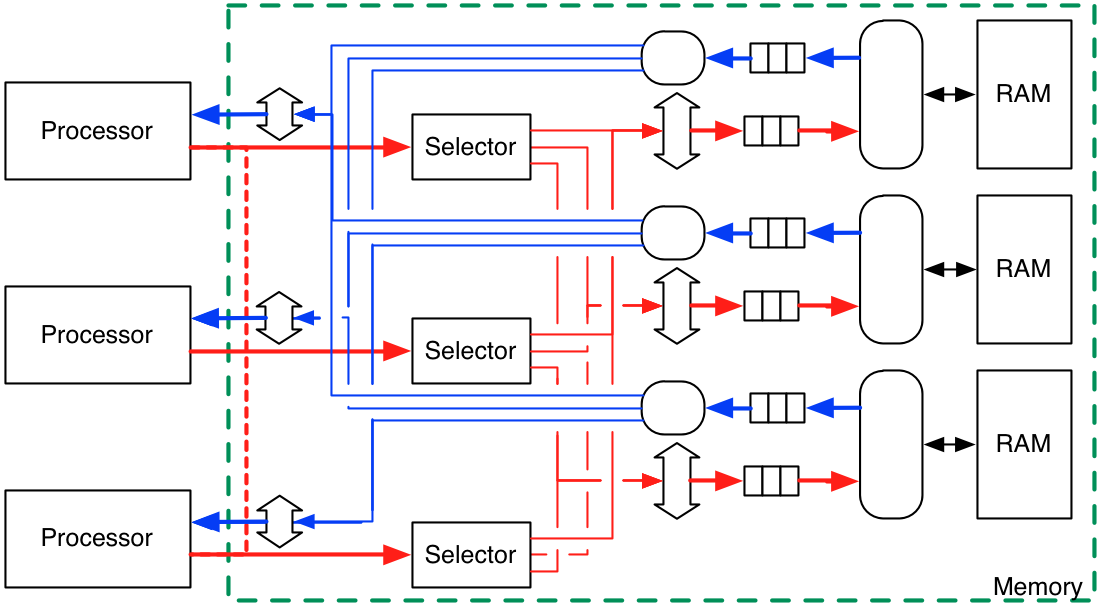}
\caption{Banked memory model in a multi-core configuration.}\label{fig:bankedmem}
\end{figure}

\begin{figure}[t]
\centering
\includegraphics[scale=.45]{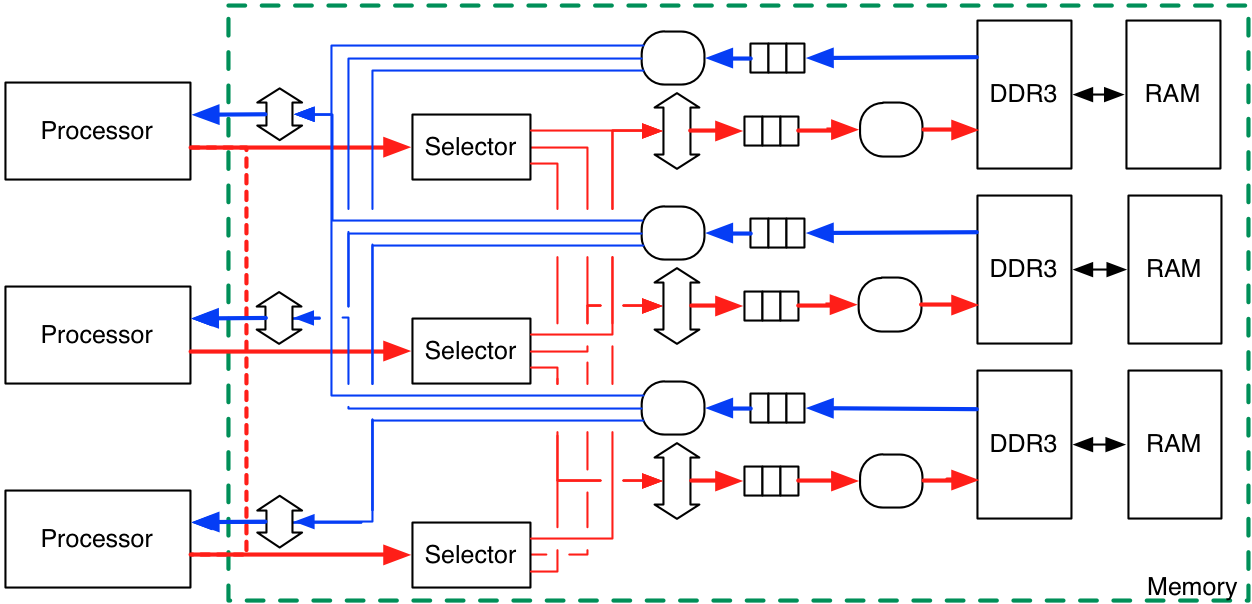}
\caption{Banked memory model with DDR3 controllers in a multi-core configuration.}\label{fig:ddrmem}
\end{figure}

Both \texttt{BankedMemory} and \texttt{DDRMemory} generalize to
multi-core configurations as shown in \cref{fig:bankedmem} and
\cref{fig:ddrmem}.

When used with multiple processors, each processor receives its own
bank selection circuit. Requests coming from a processor are then
queued to their selected bank/channel using an arbitrated service, to
appropriately simulate contention at the bank/channel.  Responses are
then delivered directly to the processor that issued them using one
arbitrator per core, as if each bank/channel controller had direct
point-to-point response links to all processors. Write requests snoop
on the L1 cache of all other processors transparently (\ie snoop
contention is not simulated).

The number of processors and the number of banks are independent, \ie
can be configured separately. If only the number of processors is
specified, the number of banks is automatically scaled to be the same.

This multi-core generalization of the two models is intended to simulate a
memory circuit in hardware with a crossbar switch or multistage
interconnection network between processors and memory banks/channels,
such as found on the Niagara chip. The hardware and energy costs of
these circuits usually prevent their practical use with more than
16-32 cores.

\subsection{Ring-based cache diffusion networks}

An ongoing research project at the University of Amsterdam aims to
design a novel memory system based on a scalable network of
caches. The latest outcome of this research is also provided as 
memory models in MGSim's library of predefined components.

There are two families of models, referred to as ``\texttt{COMA}'' and
``\texttt{ZLCOMA}'' in MGSim's source code. Their use of the acronym
``COMA'' is due to historical ties between this research and related
work on Cache-Only Memory Architectures in the
1990's~\cite{hagersten.92,dahlgren.99}. 

In the original COMA vision from the 1990's, DRAM banks in the system
are managed as a network of very large caches. The key characteristic
is that the partitioning of the address space among the memories is
not static: directories at each memory retain an index of which
addresses can be currently found at the memory. A protocol between the
memories ensures that all addresses are present \emph{somewhere} in
the network. The corresponding key feature is that the blocks of data
corresponding to the most frequently used addresses can be migrated to
the part of the system where they are most frequently used. Because of
this feature, each memory in the system is also called
\emph{attraction memory} (AM).

However, the memory architecture modeled in MGSim's library has diverged from
the original COMA concept. The reason is that it is difficult to fit
the entire DRAM storage of a system onto a single chip, and off-chip
COMA caches and directories are difficult to manage at the latency
and bandwidth scales required by on-chip parallelism. Instead,
the memory architecture designed at the University of Amsterdam still uses
static partitioning of the address space across memory banks, but
uses a distributed network of caches on chip with a similar attraction
property as a COMA: cache lines fetched once in the cache network tend
to stay within the network, so as to avoid off-chip memory
accesses as much as possible. Moreover, a cache miss in one part
of the network is resolved by fetching data from another part if it
can be found, instead of fetching from external memory. As
with a COMA, on-chip directories remember which addresses
can be currently found within the on-chip cache network. 

\begin{figure}
\centering
\subfloat[Uniform Memory Architecture (UMA)]{\includegraphics[scale=.4]{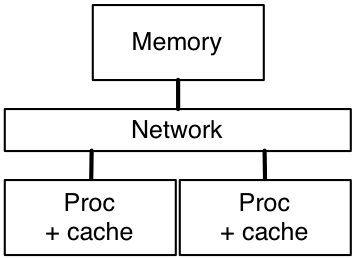}} \quad
\subfloat[Non-Uniform Memory Architecture (NUMA)]{\includegraphics[scale=.4]{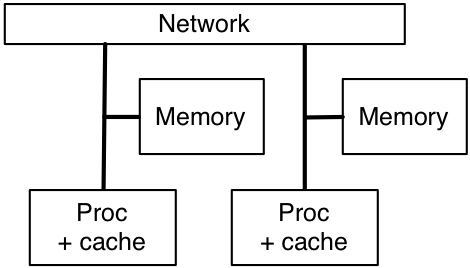}} \quad
\subfloat[Cache-Only Memory Architecture (COMA)]{\includegraphics[scale=.4]{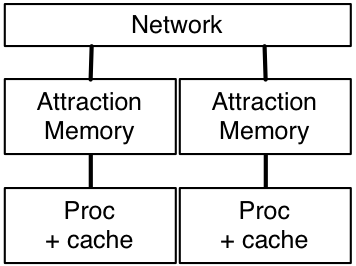}} \quad
\subfloat[Cache Diffusion Memory Architecture (CDMA)]{\includegraphics[scale=.4]{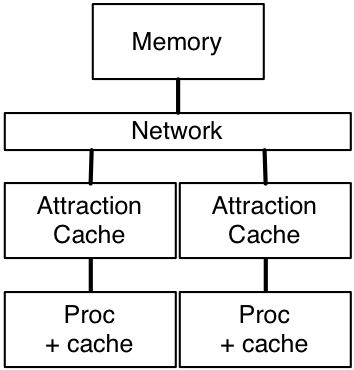}}
\caption{Overview of differences between memory architectures.}\label{fig:memarchs}
\end{figure}

Because of this difference with COMAs, the memory architecture modeled
in MGSim would be best described as a \emph{cache diffusion memory
architecture} (CDMA), based on \emph{attraction caches} (AC) and a
\emph{cache diffusion network} (CDN). We summarize the difference between
the more conventional uniform and non-uniform memory architectures,
COMAs and CDMAs in \cref{fig:memarchs}.

\subsubsection{``\texttt{COMA}'' \vs ``\texttt{ZLCOMA}''}

As mentioned initially, MGSim's library provides two CDMA models called
``\texttt{COMA}'' and ``\texttt{ZLCOMA}.''  Both use the same topology
(layout) of caches, directories and network links, presented in the
following sections.

The model ``\texttt{ZLCOMA}'' was historically implemented first, and
corresponds to the token-based, write-invalidate cache protocol
published in~\cite{zhang.07,vu.08.icamst}. Its design strives to
minimize communication across the cache network. Unfortunately, this
protocol currently relies on complex metadata updates at the caches
and directories, which makes its implementation in hardware relatively
expensive; moreover its weak consistency model violates the expectations
of shared memory programming.

These limitations have motivated the design of an alternate,
simpler, write-update cache protocol over the same component
topology. The ``\texttt{COMA}'' model uses this simpler
protocol. Because it uses write updates messages to signal memory
loads to lines duplicated in other caches (instead of requesting first
exclusivity by forcing other caches to evict their copy, as with
``\texttt{ZLCOMA}''), this protocol suffers from high bandwidth
contention for write-heavy workloads with poor cache locality.

Both protocols are research projects of the CSA group at the
University of Amsterdam. Although they are not known to be used in existing
hardware products at the time of this writing, their models
and default configuration in MGSim's library strives to reflect a
realistic potential hardware implementation in a future chip.

\subsubsection{Components in a cache diffusion network}

\begin{figure}[t]
\centering
\includegraphics[scale=.45]{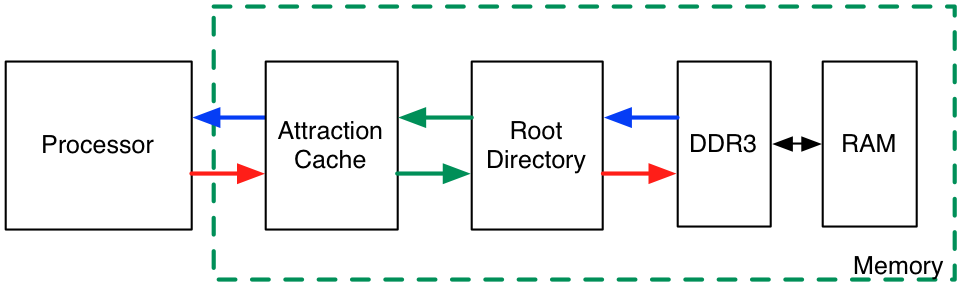}
\caption{Cache diffusion network (CDN) with only one processor, cache and
  external channel.}\label{fig:coma1}
\end{figure}

The simplest configuration of the CDMA models is depicted
in \cref{fig:coma1}. The processor is connected via its memory
interface to an attraction cache, which thus acts as its second level
 (L2) cache. The model for a single attraction cache is called
\texttt{Cache} in MGSim's framework.  The attraction cache is then
connected to a \emph{root directory} via a ring network (green links
in the picture), with which it organizes its distribution of cache
lines. The model for a root directory is called \texttt{RootDirectory}
in MGSim. The root directory is in turn connected to main memory via a
DDR3 controller, using the same DDR3 channel simulation as the
\texttt{DDRMemory} model discussed earlier in \cref{sec:bankedmem}.

At a given topology, the main configurable parameters of the
model define the structure of attraction caches: 

\begin{itemize}
\item the cache size, via the number of sets and set associativity;
\item the strategy to map memory addresses to cache set indices, using
the same selector options as in \cref{sec:bankselectors}.
\end{itemize}

Besides the cache structure, the clock frequency of the components can
be adjusted, as well as the number of buffers for each network link.
The number of directory entries in \texttt{RootDirectory} is further
automatically configured based on the cache parameters and the number
of caches in the network.

At the time of this writing, the default configuration of both the
\texttt{COMA} and \texttt{ZLCOMA} models in MGSim's library defines
4-way associative, 128KiB attraction caches with the \texttt{XORFold}
set selection strategy. Each link buffer has space for 2 requests, and
the entire cache network up to but excluding DDR3 channels is clocked
at 1GHz.

\subsubsection{CDMA topologies}

All component models in MGSim's library are accompanied with
configuration code to instantiate component models and connect
them together to form an actual architecture. To simplify
configuration, MGSim provides \emph{parameterized topologies}: 
component and architecture constructors that automatically
connect components together according to a general skeleton, based on 
few parameters, such a the desired number of components of each type.

There are two proposed network topologies for CDMAs: \emph{a single ring} and
\emph{stacked rings}. Reference configuration code is provided with the CDMA
models in the MGSim library. Each constructor is parameterized by the desired
number of components of each type, and automatically instantiate
components and connect them in ring networks.

\begin{figure}[t]
\centering
\includegraphics[scale=.45]{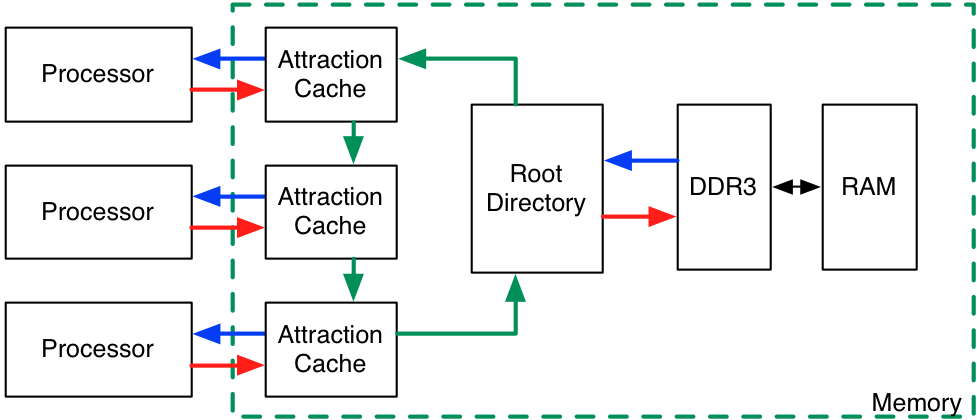}
\caption{Single-ring CDMA with three attraction caches
  and one directory.}\label{fig:coma3p1d}
\end{figure}

\begin{figure}[t]
\centering
\includegraphics[scale=.45]{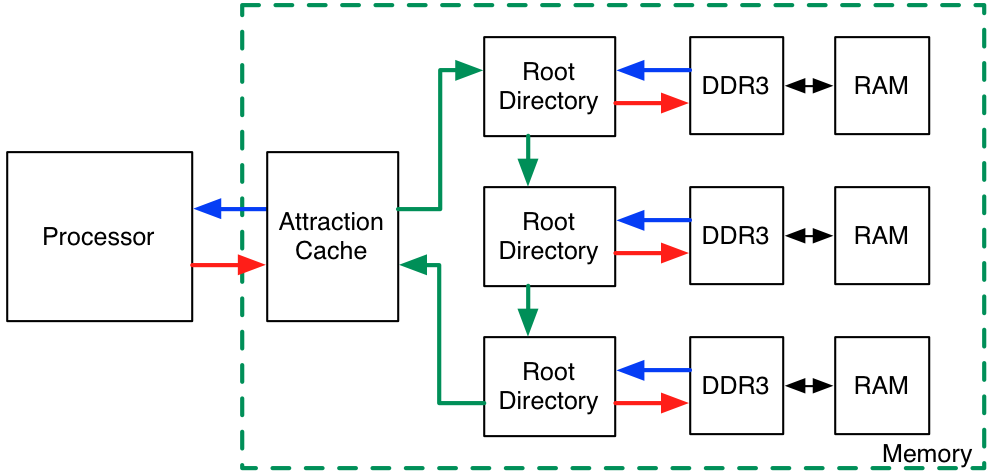}
\caption{Single-ring CDMA with one attraction cache and
  three directories.}\label{fig:coma1p1c3d}
\end{figure}

\begin{figure}[t]
\centering
\includegraphics[scale=.45]{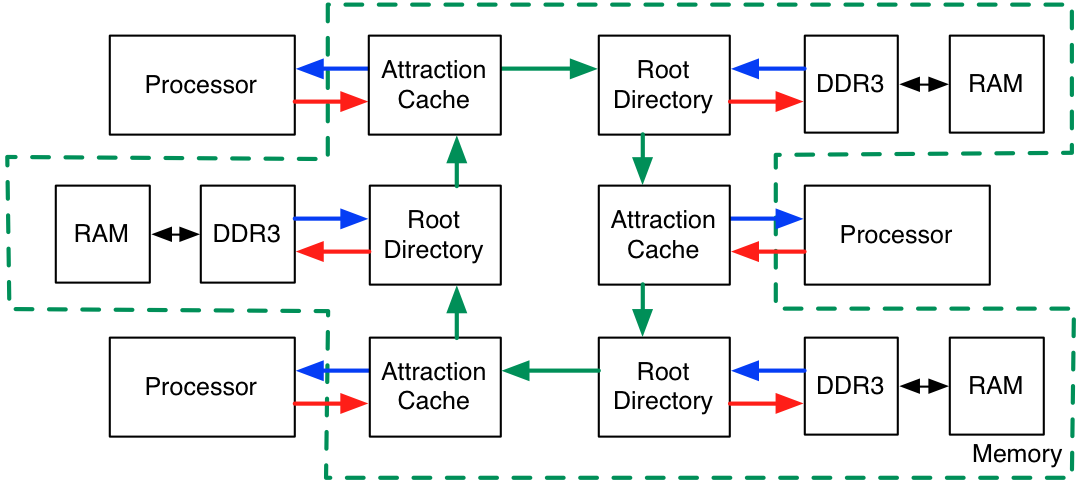}
\caption{Single-ring CDMA with three attraction caches and
  three directories.}\label{fig:coma3p3c3d}
\end{figure}

With a simple ring, both the number of attraction caches and the
number of root directories can be configured separately. As both
counts grow, the components are automatically organized as depicted in
\cref{fig:coma3p1d,fig:coma1p1c3d,fig:coma3p3c3d}.  When both multiple
attraction caches and multiple root directories are configured, the
components are automatically interleaved in the ring network, so as to
minimize the average latency of requests.

The ability to define multiple root directories provides memory
parallelism. As of this writing, the memory address space is
\emph{striped} over the available back-end memories at the granularity
of single cache lines, so as to maximize load distribution of
requests.

\begin{figure}[t]
\centering
\includegraphics[scale=.45]{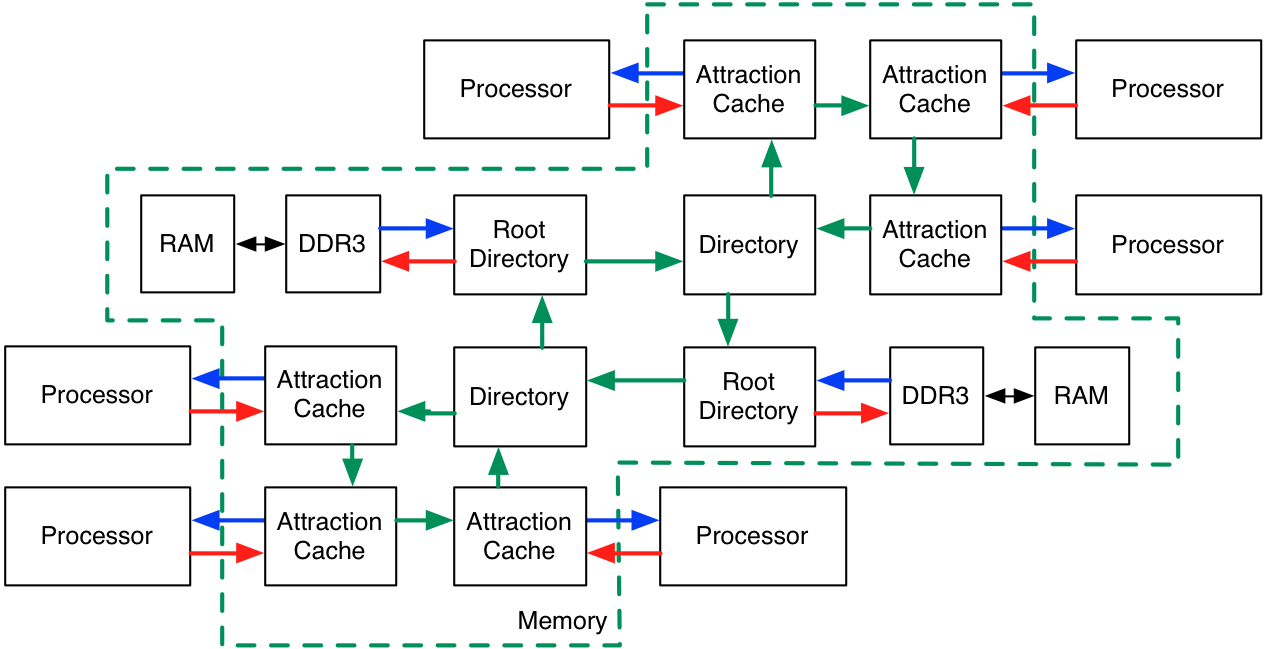}
\caption{Stacked-ring CDMA with six attraction caches and
  two root directories.}\label{fig:coma6p6c4d}
\end{figure}

The other provided topology skeleton uses stacked rings: cluster of
attraction caches are grouped into local rings, and all rings are
connected together in a top-level ring containing the root directories
to external memory. An example is given in
\cref{fig:coma6p6c4d}.

The motivation behind stacked rings is to exploit locality. Indeed, a
multi-core computation may require a lot of cache line movements
between its cores, but may be relatively independent from the memory
activity of the rest of the system. In this case, it is desirable to
segregate its network activity physically, so as to avoid influencing
negatively the bandwidth available to the rest of the system. This is
the service offered by the local rings in CDMAs: local inter-cache
traffic within the local ring stays mostly independent from the traffic from
other local rings.

With stacked rings, the connection point between local rings and the
top-level ring is a \emph{partial directory} (modeled by the
\texttt{Directory} component in MGSim), which only inventory the
lines present in the local ring. As with the single-ring topology,
partial and root directories are interleaved in the top-level ring,
and partial directories are sized automatically based on the configuration
of the local attraction caches.

\subsubsection{Shared attraction caches}

\begin{figure}[t]
\centering
\includegraphics[scale=.45]{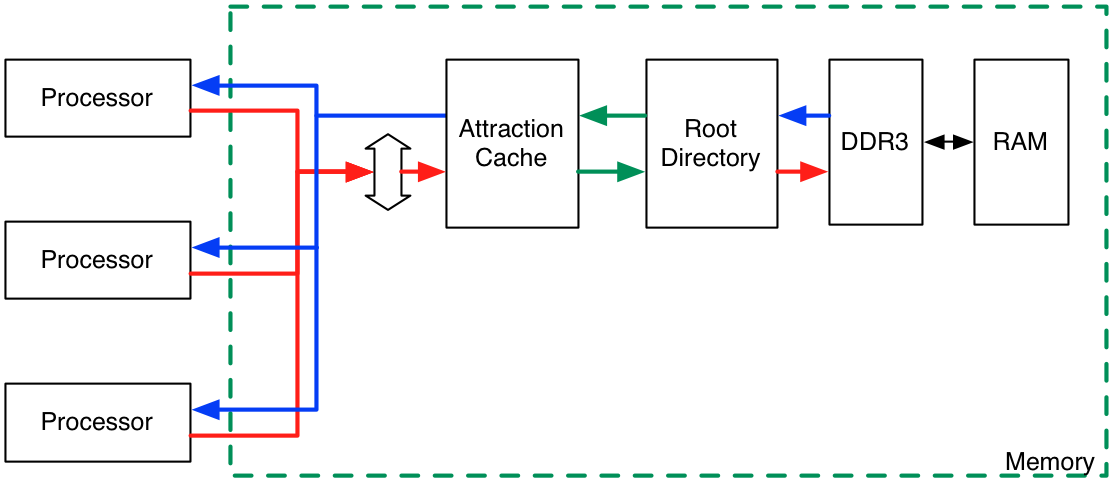}
\caption{Simple CDN with a shared attraction cache.}\label{fig:coma3p1c}
\end{figure}

Another dimension of parallelism can be obtained by sharing one
attraction cache between multiple processors, as illustrated in
\cref{fig:coma3p1c}. This type of configuration uses the same
communication model between the processors and the shared cache as the
serial memory model presented in \cref{sec:sermem}: a request bus is
used to arbitrate simultaneous requests by different processors, and
responses are served using a separate response bus.  Also, as with the
serial memory, this type of sharing is subject to contention on the
request bus.

\subsubsection{Summary of configuration parameters}

The following parameters guide the instantiation of MGSim's CDMA
models:
\begin{itemize}
\item whether to use a single ring or stacked rings;
\item the number of sets and set associativity of the attraction caches;
\item the number of root directories;
\item the total number of processors;
\item the number of processors per attraction cache;
\item the buffer sizes for the ring networks' links;
\item the clock frequency of the memory components;
\item for stacked rings, the number of attraction caches in a local ring.
\end{itemize}

The number of attraction caches, as well as the number of local rings for
stacked ring topologies, are computed automatically based on the other
configuration settings.

\subsection{Simulated I/O devices}

The MGSim library provides pseudo I/O devices to interface simulated
systems with the simulation environment. All the provided devices use
the same interface, which assumes a network-on-chip with
request/response packet-based communication (i.e. the primary
interface is not a bus). 

Additionally, the library also defines an example dedicated
network-on-chip that connects I/O devices and processor cores
together. This model separates I/O from memory operations to separate
(logical) networks; this choice was made to enable separate study of
memory-related and I/O-related traffic in the initial applications of
MGSim. The overall design of the I/O system and its motivations have been published in~\cite{poss.12.rapido}.

\subsubsection{Real time clock}

A \texttt{RTC} pseudo-device is provided; it gives
access to real time to simulated systems. Two operation modes
are supported:
\begin{itemize}
\item read requests can \emph{query} the current time;
\item write requests can \emph{configure asynchronous timer events} which
are delivered on-time over the I/O network. 
\end{itemize}

The latter facility can be exploited at the processor core interface
to trigger periodic processing.

\subsubsection{Console pseudo-devices}

Three pseudo-devices are provided to interface with the user.

\texttt{UART} implements a serial line, using a subset of the standard
NS/PC16650\footnote{\url{http://www.national.com/ds/PC/PC16550D.pdf}}
programming interface.  On the simulation host, the serial line can be
connected either to the terminal, to a file, or to a Unix FIFO towards
other processes.

\texttt{LCD} implements a character matrix display. It has
   two programming modes: a memory array where each index corresponds
   to a character on display; and a serial byte-oriented interface which
   automatically implements newlines and scrolling. On the simulation host
   the display is rendered on the operator's text terminal.

\texttt{Display} implements a pixel-oriented display. It 
   supports its own video memory of configurable size, which serves
   as frame buffer. It also supports palette (indexed) and ``true'' (RGB) color modes.
   On the simulation host the display is rendered using a graphical display.

\subsubsection{Interface to the host file system}

The pseudo-device \texttt{RPC} implements a virtual file system
server. It is operated using a remote procedure call protocol. Using
I/O read and write requests to the pseudo-device, the simulated system
can perform calls to standard filesystem operations in the host
environment: \texttt{open}, \texttt{read}, \texttt{write}, etc.

This facility was designed to enable benchmarks that need large
input/output data files to feed compute-intensive algorithms, where
the focus was to evaluate the computation part. By providing direct
access through a RPC interface, the simulation can bypass the overhead
of full operating system simulation.

\subsubsection{Read-only memory with DMA controller}

The pseudo-device \texttt{ActiveROM} implements a combination
of a read-only memory (ROM) with a DMA controller. 

The main use of this device is to provide access to the content of
specific files on the simulation host, for example the executable code
of a benchmark program. Two additional options are also available: automatic
initialization to the set of Unix environment variables, and to the list
of command-line arguments; this facilitates the execution of multiple simulations
across ranges of software parameters provided externally.

As expected, the device returns the ROM data upon read requests. Write
requests configure and trigger DMA operation: the embedded DMA controller
is able to asynchronously send regions of the ROM over the network-on-chip to
another node on the network. 

This facility was designed to automate loading the initial
bootstrapping software code into RAM at the start of simulations.

\subsubsection{System management controller}

The pseudo-device \texttt{SMC} implements the initialization sequence for
the simulated system and a device discovery protocol.

The device discovery is implemented using a small ROM which is
automatically populated with a listing of which nodes (address,
component type, model) are connected to the I/O network.

Upon system initialization, the \texttt{SMC}  components become
active and perform the following actions:
\begin{enumerate}
\item they trigger a peer \texttt{ActiveROM} device to send the
  bootable executable code to RAM via the cache controller of
  one of the processor core(s) connected to their I/O network;
\item once the transfer is completed, they trigger the boot signal of
  the processor core(s). As part of the
  boot signal they also inform the processor core(s) of their own
  address on the I/O network. This way the software running within the
  simulation can dynamically discover which devices are actually
  configured.
\end{enumerate}

\subsubsection{Example I/O interconnect}

MGSim's library provide a model of I/O interconnect called
\texttt{NullIO}. This implements a fully connected network between all
nodes, with a node-to-node latency of 1 network cycle (the network's
clock frequency is individually configurable). The communication
latency is furthermore independent of the packet size, i.e. the network appears
to have infinite bandwidth.

While this component is not a realistic model of most I/O interconnect
in use in real-world systems, it demonstrates the I/O protocol and
enables a functional full-system emulation environment when the I/O
costs are not the focus of measurements. Besides, its implementation is
simple and can thus be used as an example for future work.

\subsubsection{Hardware/software interface}

As explained above, MGSim's standard I/O subsystem assumes a
packet-oriented network between pseudo I/O devices and processor
cores. The I/O protocol supports messages of the following types:

\begin{tabular}{ll}
Message & Arguments  \\
\hline
read request & address, size \\
read response & payload \\
write request & address, size, payload \\
broadcast\footnotemark & channel, payload \\
\end{tabular}

\footnotetext{The broadcast message type is used to signal asynchronous
(unexpected) events. The difference with the other message types is that
broadcasts are delivered to all nodes in the network. Broadcasts are
called \texttt{Notification} in MGSim's source code.} 

\begin{figure}[t]
\centering
\includegraphics[scale=.45]{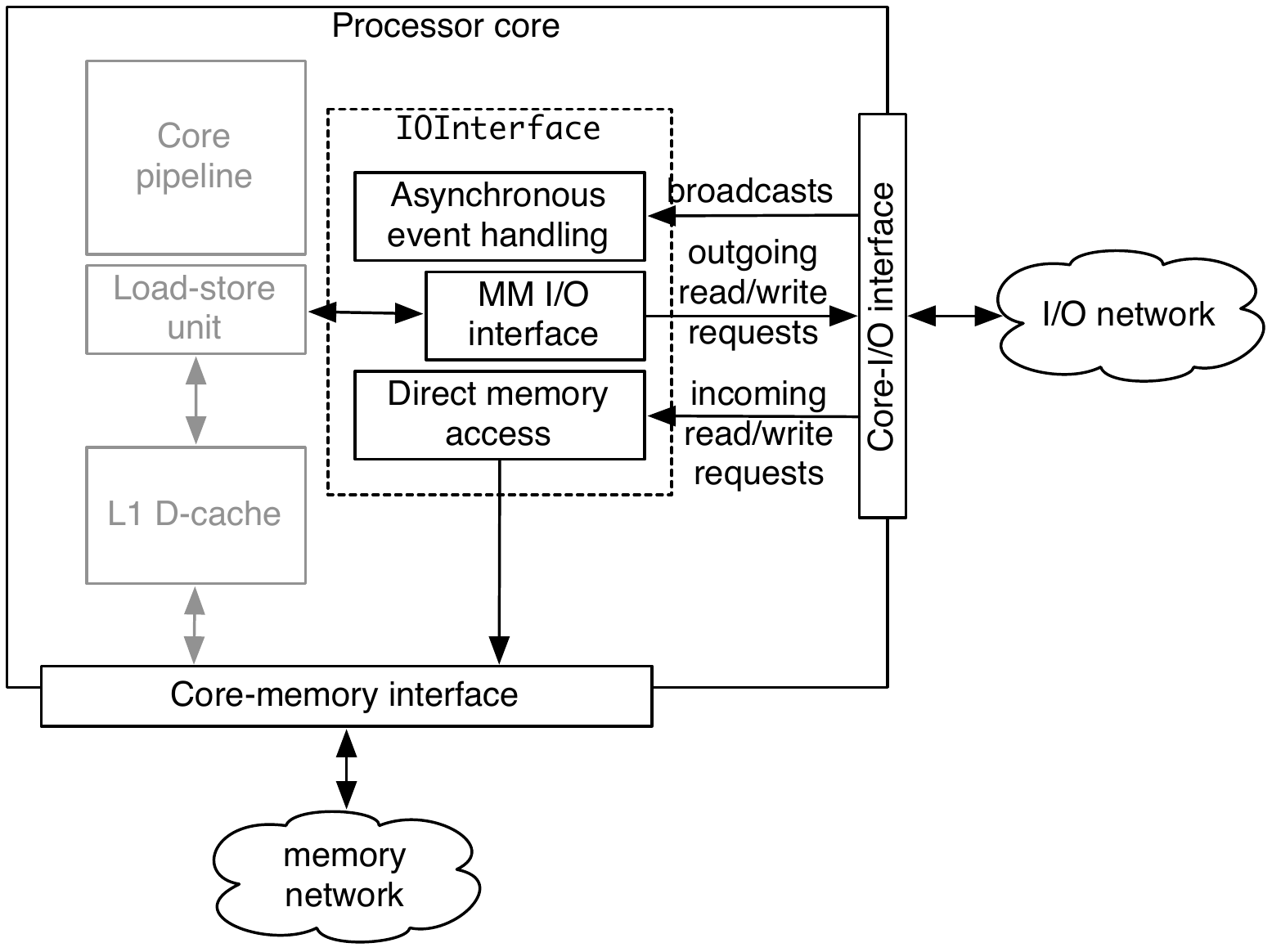}
\caption{Example use of \texttt{IOInterface} in a processor core model.}\label{fig:mmio}
\end{figure}

In contrast, most processor cores in use today assume a bus interface
to a memory system (address/data lines) and interrupt lines.  To
connect cores to the I/O network, MGSim's library thus proposes a component
\texttt{IOInterface} that can be reused in core models. This component
provides the following services:

\begin{itemize}
\item it can translate memory loads and stores issued by software
  running on the core to outgoing I/O read/write requests,
  i.e. it implements uncached memory-mapped I/O;
\item it can translate incoming I/O broadcast messages to the
  core's preferred mechanism for asynchronous signalling
  (e.g. interrupts);
\item it can translate \emph{incoming} I/O read/write requests to
   direct accesses to the core's memory interface, as a form
   of direct memory data injection from the I/O network.
\end{itemize}

This is illustraged in \cref{fig:mmio}.
MGSim's default D-RISC \texttt{Processor} model uses
\texttt{IOInterface} to implement memory-mapped I/O, and reports I/O
broadcasts via dataflow synchronizers. This mechanism is based
on the techniques published in~\cite{hicks.10.samos}.

\section{Monitoring and measurements}\label{sec:mon}

The MGSim provides three main facilities to track the simulation process:
\begin{itemize}
\item \emph{synchronous event traces}, when enabled, report \emph{all} simulation events
pertaining to a specific category. This is intended for troubleshooting or inspecting the detailed
cycle-to-cycle behavior of components;
\item \emph{asynchronous monitoring}, when enabled, asynchronously \emph{samples} the
state of the simulator at regular time intervals. This mechanism is faster than synchronous
event traces, but also less accurate. It is intended for collecting statistical
information over larger time scales;
\item \emph{performance counters} create a bridge between the software running
on the simulated system and the simulation environment: they enable programs to
read the simulation state as values and combine them with computation results. This
mechanism is intended to capture time and resource usage of individual software
components in larger benchmark programs.
\end{itemize}

These facilities are further described in the following sections.

\subsection{Synchronous event traces}

\begin{lstlisting}[language=c++,float,caption={Behavior of the D-RISC model for branches.},label=lst:bra]
if (target != next) {
    DebugFlowWrite("F%u/T%u branch %s", fid, tid, pc_str);
    COMMIT {
        m_output.pc   = target;
        m_output.swch = true;
    }
    return PIPE_FLUSH;
}
\end{lstlisting}

Component cycle handlers, or the services they invoke, can use an MGSim API
to report simulation events. For example, the pipeline execute stage
of the provided D-RISC core model uses the code in \cref{lst:bra} on
its control path for branch instructions. \texttt{DebugFlowWrite}
is a macro which only formats and prints the message defined by its arguments
if the corresponding trace type (here ``flow'') is enabled. 
The overhead to format the event text is avoided when tracing is disabled.

\begin{table}
\begin{tabular}{>{\ttfamily}ll}
{\normalfont Type} & Description \\
\hline
mem & Memory load and store instructions \\
pipe & Events reporting pipeline activity \\
reg & Accesses to a core's register file \\
fpu & Events in the floating-point unit \\
flow & Branch instructions (and thread creation for D-RISC) \\
io & I/O operations \\
ionet & Messages on the I/O network \\
deadlock & Events reporting a process stall \\
net & (D-RISC only) Events on the delegation/link network \\
sim & General simulation events (default) \\
\end{tabular}
\caption{Synchronous trace types.}\label{tab:traces}
\end{table}

The trace types currently defined are listed in \cref{tab:traces}.
Individual traces can be enabled or disabled interactively
at MGSim's command prompt by the user. 

Each line in the trace contains the following items, in this order:
\begin{enumerate}
\item the simulation's master cycle counter;
\item the name of the originating component;
\item the name of the process in the component whose handler produced the event;
\item a single-character key identifying the event category;
\item the text of the event.
\end{enumerate}

\begin{lstlisting}[basicstyle={\ttfamily\tiny},columns=fullflexible,float,caption={Example synchronous flow trace.},label=lst:extrace]
[01414660:cpu0.pipeline.execute]  (cpu0.pipeline:pipeline) f F0/T0(2063597568) <main+0x24> branch <printf+0x8>
[01415332:cpu0.pipeline.execute]  (cpu0.pipeline:pipeline) f F0/T0(2063597568) <printf+0x64> branch <vfprintf+0x8>
[01415508:cpu0.pipeline.execute]  (cpu0.pipeline:pipeline) f F0/T0(2063597568) <vfprintf+0x24> create CPU1/F31 <__slFfmta_t_vfprintf>
[01415620:cpu0.pipeline.execute]  (cpu0.pipeline:pipeline) f F0/T0(2063597568) <vfprintf+0x38> sync CPU1/F31
[01415664:cpu1.pipeline.execute]  (cpu1.pipeline:pipeline) f F31/T0(0) <__slFfmta_t_vfprintf+0x28> branch <__vfprintf+0x8>
[01416996:cpu1.pipeline.execute]  (cpu1.pipeline:pipeline) f F31/T0(0) <__vfprintf+0x84> branch <__vfprintf+0xa0>
[...]
h[01418700:cpu1.pipeline.execute] (cpu1.pipeline:pipeline) f F31/T0(0) <__vfprintf+0x1288> branch <__vfprintf+0x125c>
e[01418792:cpu1.pipeline.execute] (cpu1.pipeline:pipeline) f F31/T0(0) <__vfprintf+0x1288> branch <__vfprintf+0x125c>
l[01418884:cpu1.pipeline.execute] (cpu1.pipeline:pipeline) f F31/T0(0) <__vfprintf+0x1288> branch <__vfprintf+0x125c>
l[01418976:cpu1.pipeline.execute] (cpu1.pipeline:pipeline) f F31/T0(0) <__vfprintf+0x1288> branch <__vfprintf+0x125c>
o[01419068:cpu1.pipeline.execute] (cpu1.pipeline:pipeline) f F31/T0(0) <__vfprintf+0x1288> branch <__vfprintf+0x125c>
 [01419164:cpu1.pipeline.execute] (cpu1.pipeline:pipeline) f F31/T0(0) <__vfprintf+0x128c> branch <__vfprintf+0x1244>
[...]
[01420860:cpu1.pipeline.execute]  (cpu1.pipeline:pipeline) f F31/T0(0) <__vfprintf+0x1d44> branch <strlen+0x8>
[01420936:cpu1.pipeline.execute]  (cpu1.pipeline:pipeline) f F31/T0(0) <strlen+0x20> branch <strlen+0x34>
[...]
[01421396:cpu1.pipeline.execute]  (cpu1.pipeline:pipeline) f F31/T0(0) <strlen+0xc4> branch <__vfprintf+0x1d48>
[01421440:cpu1.pipeline.execute]  (cpu1.pipeline:pipeline) f F31/T0(0) <__vfprintf+0x1d58> branch <__vfprintf+0x1bd4>
[...]
w[01424112:cpu1.pipeline.execute] (cpu1.pipeline:pipeline) f F31/T0(0) <__vfprintf+0xc88> branch <__vfprintf+0xc5c>
o[01424204:cpu1.pipeline.execute] (cpu1.pipeline:pipeline) f F31/T0(0) <__vfprintf+0xc88> branch <__vfprintf+0xc5c>
r[01424296:cpu1.pipeline.execute] (cpu1.pipeline:pipeline) f F31/T0(0) <__vfprintf+0xc88> branch <__vfprintf+0xc5c>
l[01424388:cpu1.pipeline.execute] (cpu1.pipeline:pipeline) f F31/T0(0) <__vfprintf+0xc88> branch <__vfprintf+0xc5c>
d[01424484:cpu1.pipeline.execute] (cpu1.pipeline:pipeline) f F31/T0(0) <__vfprintf+0xc8c> branch <__vfprintf+0x6ec>
[01425088:cpu1.pipeline.execute]  (cpu1.pipeline:pipeline) f F31/T0(0) <__vfprintf+0x704> branch <__vfprintf+0x6c>
[...]
[01426532:cpu1.pipeline.execute]  (cpu1.pipeline:pipeline) f F31/T0(0) <__vfprintf+0xd94> branch <__slFfmta_t_vfprintf+0x2c>
[01426660:cpu0.pipeline.execute]  (cpu0.pipeline:pipeline) f F0/T0(2063597568) <vfprintf+0x50> detach CPU1/F31
[01426668:cpu0.pipeline.execute]  (cpu0.pipeline:pipeline) f F0/T0(2063597568) <vfprintf+0x58> branch <printf+0x68>
[01426744:cpu0.pipeline.execute]  (cpu0.pipeline:pipeline) f F0/T0(2063597568) <printf+0x78> branch <main+0x28>
[01427080:cpu0.pipeline.execute]  (cpu0.pipeline:pipeline) f F0/T0(2063597568) <main+0x40> branch <_start+0x48>
\end{lstlisting}

An example trace is given in \cref{lst:extrace}. This was produced by
compiling the C program ``\texttt{int main(void)\{ printf("hello
  \%s\textbackslash{}n", "world"); return 0; \}}'', using
the toolchain presented in~\cite{poss.12}, setting a breakpoint on
  ``\texttt{main},'' then enabling the ``\texttt{flow}'' trace after
  the breakpoint was reached. As the example illustrates, this event
  trace reveals the run-time call tree of the program and how much
time is spent in each function.

\begin{figure}[t]
\centering
\includegraphics[width=\linewidth]{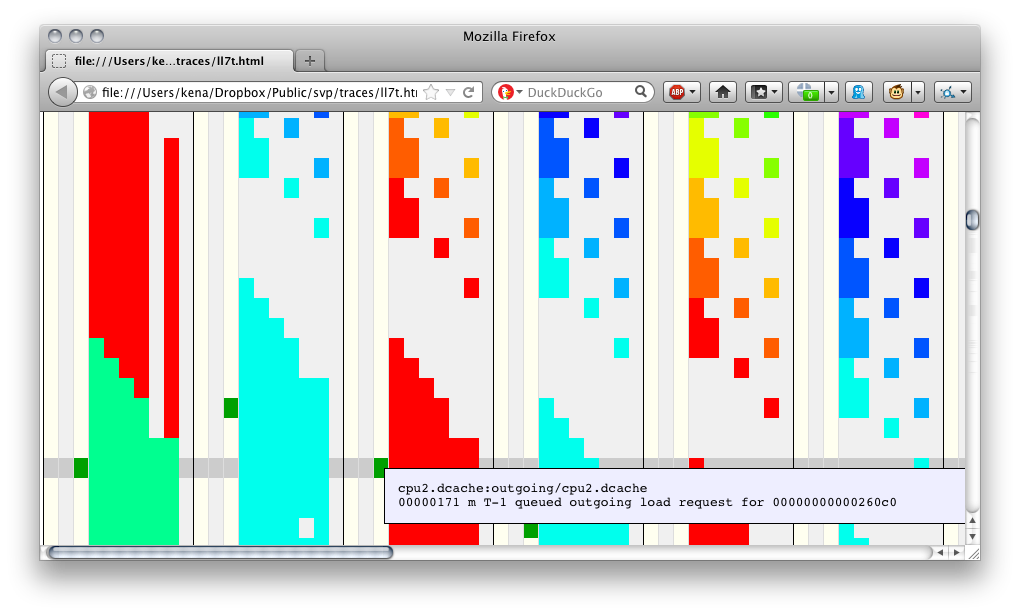}
\caption{Example use of the \texttt{viewlog} utility with synchronous event traces.}\label{fig:viewlog}
\end{figure}

Traces can be printed on the terminal, captured to files or piped
through external utilities. MGSim provides an example utility called
\texttt{viewlog} that converts event traces to HTML files suitable for
graphical visualizations in a web browser. An example output is
provided in \cref{fig:viewlog}. In this example, a 8-core model is
running a parallelized implementation of the equation of state
fragment found as loop 7 of the Livermore benchmark
suite~\cite{mcmahon.86}. The output of \texttt{viewlog} produces one
column per component and one row per cycle. Within one column,
different colors are used for different hardware threads in the D-RISC
core model. The browser window is centered in the start of the
benchmark's data-parallel operation; the cursor is hovering at the
intersection between cycle 171 and the L1 D-Cache of core 2, and a
pop-up label shows a memory event occurring at that location.

\subsection{Asynchronous monitoring}

\begin{figure}[t]
\centering
\includegraphics[scale=.45]{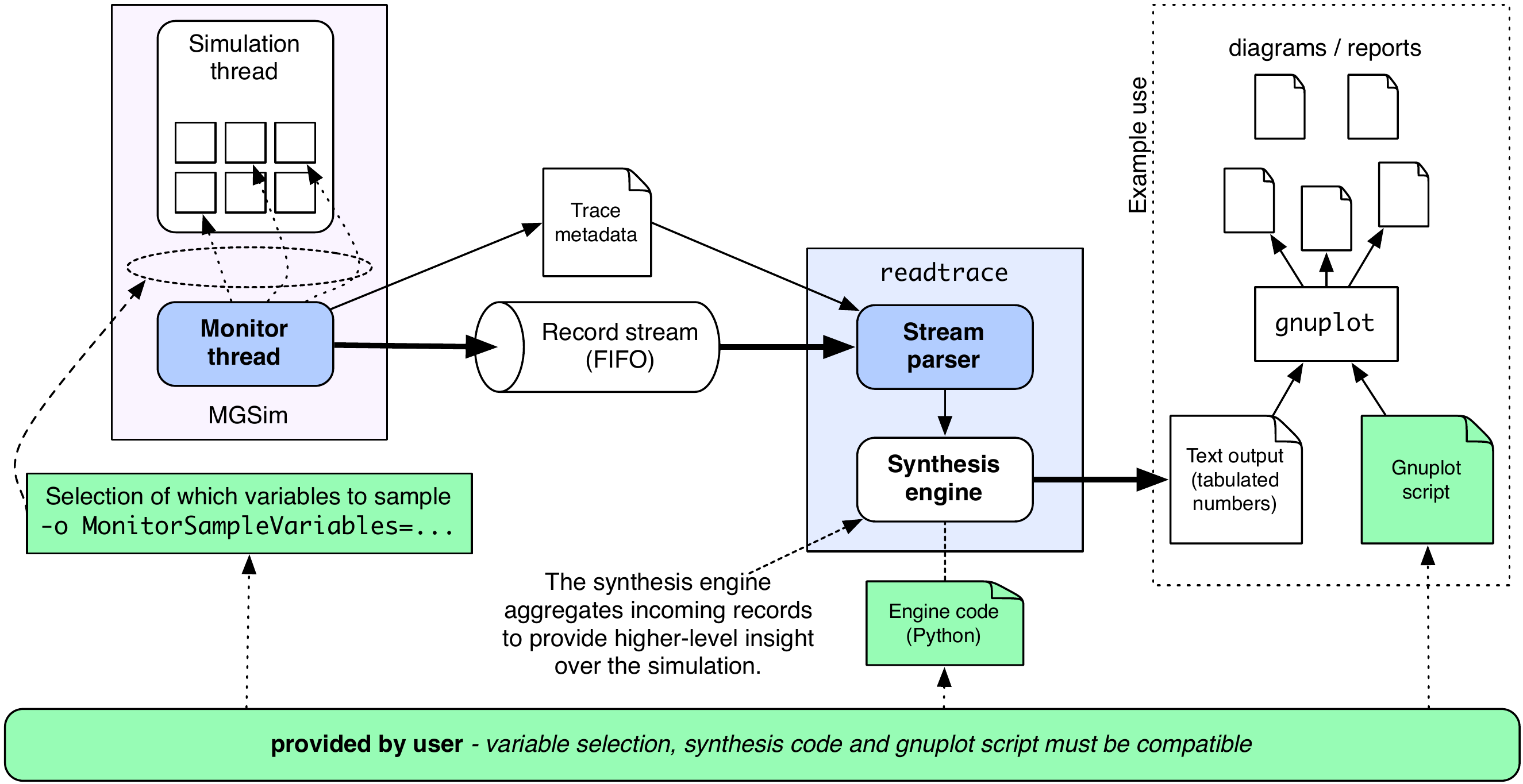}
\caption{Architecture and use case for asynchronous monitoring in MGSim.}\label{fig:monitor}
\end{figure}

Asynchronous monitoring is intended to capture the evolution over time
of semi-continuous variables in the simulation model. Its architecture
is described in \cref{fig:monitor}: a monitor thread runs concurrently
with the simulation thread, and repeatedly samples a set of selected
variables as binary records to an output stream. The sampling rate is
configurable.

The stream can be redirected to a file, or piped through an external
utility. MGSim provides a generic utility \texttt{readtrace} which can
transform the binary format to text, suitable for plotting using
e.g. GNUPlot. \texttt{readtrace} can also be used to reduce and
aggregate the monitor samples.

\begin{figure}[t]
\centering
\includegraphics[width=\linewidth]{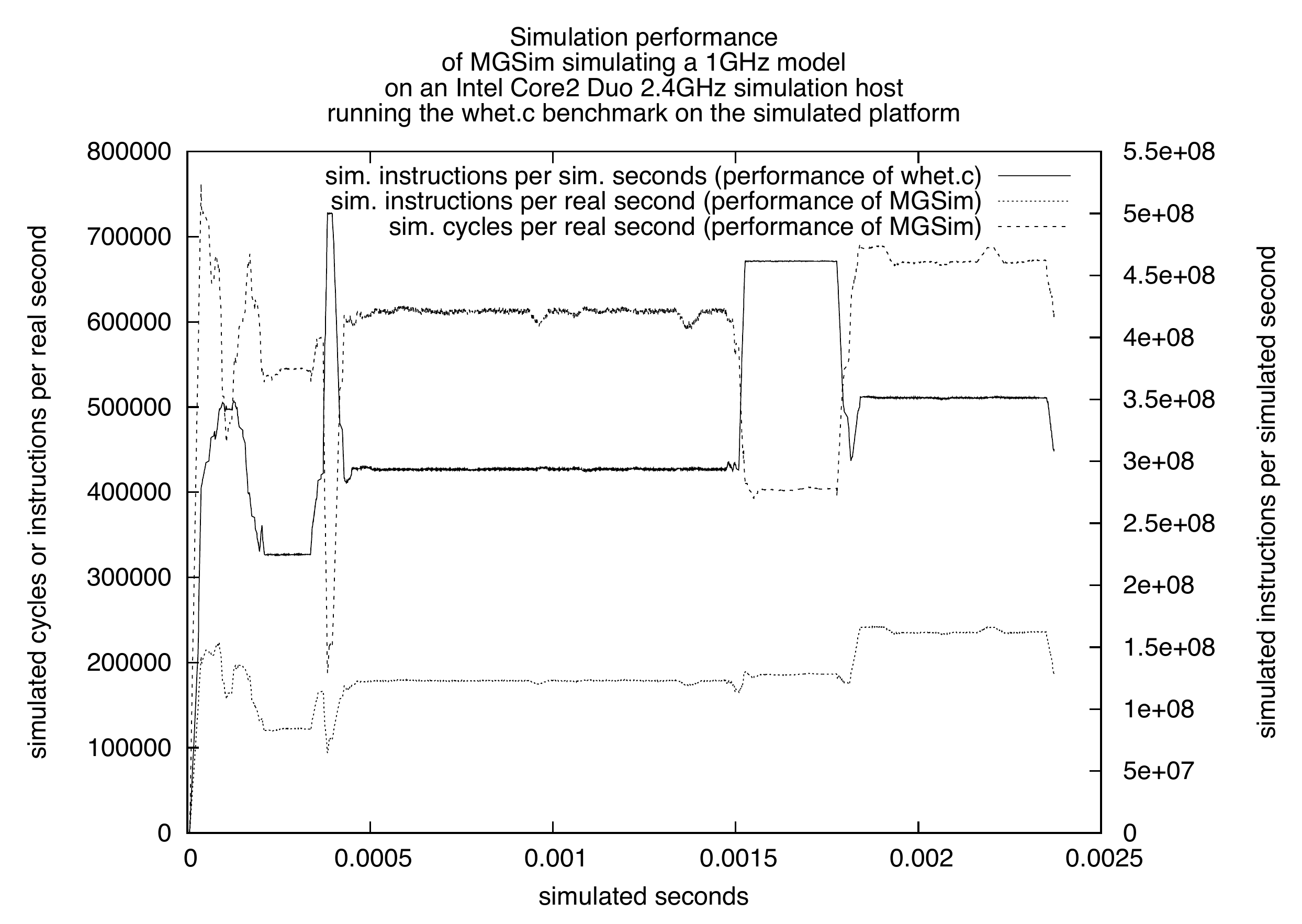}
\caption{Example application of asynchronous monitoring.}\label{fig:readtrace}
\end{figure}

An example use case is illustrated in \cref{fig:readtrace}. In this
example, a 16-core simulation model was configured to run the
classical Whetstone
benchmark\footnote{\url{http://www.netlib.org/benchmark/whetstone.c}}
(a sequential program) using an Apple MacBook Pro as simulation
host. The monitor thread was configured to sample the simulation cycle
counter and the counter for the number of instructions executed in the
cores' pipeline.  The sampling rate was configured to 1000
samples/s. The execution of \texttt{whet.c} lasted for approximately
4.22 real seconds, little over 2.3ms of simulated time; it ran about
3.1M instructions.  During this time, the monitor thread produced 3705
samples at an approximated effective rate of 876 samples/s. The
\texttt{readtrace} utility was subsequently used in conjunction with
GNUplot to produce the figure. 

As the figure shows, for this model and this host MGSim runs
approximately 200K instructions per real second (KIPS) in general, or
150-200x slower than an equivalent hardware implementation. This perspective gives
a better view of the simulation speed than the naive estimation based
on the final counts, which indicate $3.1\times 10^6/4.22$ or about
740KIPS for this particular program.

\subsection{Performance counters}

Next to the monitoring facilities provided above, which are fully ``external'' and
invisible to the simulated platform, MGSim's library also provides an interface
to enable performance introspection by simulated programs.

As a first step, MGSim's standard core model support the standard ISA
features to expose performance counters where they exist. For example,
its Alpha-based D-RISC model supports Alpha's \texttt{rpcc}
instruction to read the core's cycle counter.

Moreover, MGSim's library provides a component called
\texttt{PerfCounters}.  This can be used by any core model at its
memory interface to redirect loads to a specific range of addresses to
performance counters. When this is configured, a program can
introspect the progress of the simulation by reading from fixed memory
ranges. At the time of this writing \texttt{PerfCounters} provides
access to counters for the number of clock cycles, executed
instructions and issued memory requests, between others.

This mechanism was implemented to enable precise reporting of MGSim
statistics from the start to end points of a computation expressed in
source code (e.g. C). It is not intended as a realistic model of
performance counter interfaces found in real hardware.

\section{Shortcomings and possible future work}\label{ssec:fut}

We have presented so far the state of MGSim's framework and its
standard component library at the start of 2013. Our experience
using MGSim for architecture research and education has revealed
two serious shortcomings.

The first is the common occurrence of implementation or design errors
when implementing a new model in MGSim. The most common error is the
definition of deadlocking circuits due to circular
dependencies. Although the component model exposes all dependencies
between buffers and processes, the MGSim framework is not yet able to
analyze and detect circular dependencies automatically.  The
implementation of such a detection mechanism would significantly reduce
the time required to troubleshoot modeling errors.

The second shortcoming is the lack of a facility to checkpoint/restore
the entire simulation state. When a failure occurs, the only mechanism
available to-date to reproduce the issue is to re-play the entire
execution scenario since the start of the simulation. If the program
further uses I/O, an exact re-execution is simply impossible. This
becomes an issue particularly when troubleshooting long-running
software within the simulated platform.  Mechanisms to serialize and
de-serialize the simulation state, similarly to the ``freeze'' feature
of virtual machines, would greatly increase the suitability of MGSim
as a sandbox environment to troubleshoot simulated software.

Next to these shortcomings, the question arose of what to do about the
similarities between MGSim and Gem5, discussed previously in
\cref{sec:related}. Despite the different project goals, the overlap
between the technical approaches is striking; in particular, the
inter-component interfaces, component granularity and configuration
facilities are intriguingly similar between the two projects. This
raises two opportunities. The first is to investigate whether MGSim's
library of memory models could be reused with Gem5, which is somewhat
still lacking in this regard. The other is to determine whether Gem5's
core models could be reused with MGSim, to provide increased platform
compatibility to programs running on the simulated platform.

\section{Summary and conclusions}\label{sec:conc}

We have presented MGSim, an open source framework and component
library to simulate many-core processors. MGSim's framework is written
in C++ and implements a highly configurable, discrete-event,
multi-clock simulation engine. Its library of components provides a
versatile hardware multithreaded in-order RISC core supporting
multiple ISAs, multiple memory interconnects, and an I/O subsystem
which enables full-system emulations. Its comprehensive inspection
and monitoring facilities make it suitable for both architecture
research and education.

MGSim is currently used at the University of Amsterdam and its
partners. Its applications include scientific research on the Microgrid
architecture~\cite{poss.12.dsd} and general graduate-level education on
processor, cache and memory architectures.  Performance-wise, MGSim is
known to run models containing thousands of components at 100-1000KIPS
on conventional desktop-grade hardware.

MGSim is similar to Gem5~\cite{binkert.06.micro}, another C++-based
framework for discrete-event, component-based multi-core simulations. The two
frameworks run with comparable performance. Where Gem5 focuses on
compatibility with real hardware and intra-core accuracy on models
with few cores, MGSim focuses on implementation simplicity and
accuracy with large many-core models.

\section*{Acknowledgements}
\addcontentsline{toc}{section}{Acknowledgements}

MGSim's framework and main component models were originally
designed and implemented by Mike Lankamp. MGSim is currently under
stewardship of the authors of this article.  The development of MGSim
so far was funded by the Dutch government via the project NWO
Microgrids, the European Union under grant numbers FP7-215216
(Apple-CORE) and FP7-248828 (ADVANCE), the University of Amsterdam,
and grants by the China Scholarship Council.

The authors would like to thank Carl Joslin, Michiel W. van Tol,
Thomas Bernard and Andrei Matei for their thorough evaluation and
testing of the MGSim tool chain during its inception, as well as Zhang
Li for his contribution of the CDMA model ``\texttt{ZLCOMA}.''

\newcommand{\etalchar}[1]{#1} 
\addcontentsline{toc}{section}{References}
\bibliographystyle{is-plainurl}
\bibliography{doc}

\begin{thebibliography}{10}
\ifx \showCODEN  \undefined \def \showCODEN #1{CODEN #1}  \fi
\ifx \showISBN   \undefined \def \showISBN  #1{ISBN #1}   \fi
\ifx \showISSN   \undefined \def \showISSN  #1{ISSN #1}   \fi
\ifx \showLCCN   \undefined \def \showLCCN  #1{LCCN #1}   \fi
\ifx \showPRICE  \undefined \def \showPRICE #1{#1}        \fi
\ifx \showURL    \undefined \def \showURL {URL }          \fi
\ifx \path       \undefined \input path.sty               \fi
\ifx \ifshowURL \undefined
     \newif \ifshowURL
     \showURLtrue
\fi

\bibitem{bernard.08.samos}
T.~Bernard, K.~Bousias, L.~Guang, C.~R. Jesshope, M.~Lankamp, M.~W. van Tol,
  and L.~Zhang.
\newblock {A general model of concurrency and its implementation as many-core
  dynamic RISC processors}.
\newblock In W.~Najjar and H.~Blume, editors, {\em Proc. Intl. Conf. on
  Embedded Computer Systems: Architecture, Modeling and Simulation (IC-SAMOS
  2008)}, pages 1--9. IEEE, Samos, Greece, July 2008.
\newblock \showISBN{978-1-4244-1985-2}.

\bibitem{binkert.06.micro}
Nathan~L. Binkert, Ronald~G. Dreslinski, Lisa~R. Hsu, Kevin~T. Lim, Ali~G.
  Saidi, and Steven~K. Reinhardt.
\newblock The {M5} simulator: Modeling networked systems.
\newblock {\em IEEE Micro}, 26:\penalty0 52--60, 2006.
\newblock \showISSN{0272-1732}.
\newblock \href {http://dx.doi.org/10.1109/MM.2006.82}
  {\path{doi:10.1109/MM.2006.82}}.

\bibitem{bolychevsky.96.ieee}
A.~Bolychevsky, C.R. Jesshope, and V.B. Muchnick.
\newblock Dynamic scheduling in {RISC} architectures.
\newblock {\em IEE Proceedings - Computers and Digital Techniques},
  143\penalty0 (5):\penalty0 309--317, September 1996.
\newblock \showISSN{1350-2387}.
\newblock \href {http://dx.doi.org/10.1049/ip-cdt:19960788}
  {\path{doi:10.1049/ip-cdt:19960788}}.

\bibitem{dahlgren.99}
F.~Dahlgren and J.~Torrellas.
\newblock Cache-only memory architectures.
\newblock {\em Computer}, 32\penalty0 (6):\penalty0 72--79, June 1999.
\newblock \showISSN{0018-9162}.
\newblock \href {http://dx.doi.org/10.1109/2.769448}
  {\path{doi:10.1109/2.769448}}.

\bibitem{vu.08.icamst}
T.~D.Vu, L.~Zhang, and C.~R. Jesshope.
\newblock {The verification of the on-chip COMA cache coherence protocol}.
\newblock In {\em International Conference on Algebraic Methodology and
  Software Technology}, pages 413--429, 2008.
\newblock \showISBN{978-3-540-79979-5}.

\bibitem{gonzalez.96.isc}
Antonio Gonz\'{a}lez, Mateo Valero, Nigel Topham, and Joan~M. Parcerisa.
\newblock Eliminating cache conflict misses through {XOR}-based placement
  functions.
\newblock In {\em Proc. 11th International Conference on Supercomputing
  (ICS'97)}, pages 76--83. ACM, New York, NY, USA, 1997.
\newblock \showISBN{0-89791-902-5}.
\newblock \href {http://dx.doi.org/10.1145/263580.263599}
  {\path{doi:10.1145/263580.263599}}.

\bibitem{hagersten.92}
E.~Hagersten, A.~Landin, and S.~Haridi.
\newblock Ddm-a cache-only memory architecture.
\newblock {\em Computer}, 25\penalty0 (9):\penalty0 44--54, September 1992.
\newblock \showISSN{0018-9162}.
\newblock \href {http://dx.doi.org/10.1109/2.156381}
  {\path{doi:10.1109/2.156381}}.

\bibitem{hicks.10.samos}
Michael~A. Hicks, Michiel~W. van Tol, and Chris~R. Jesshope.
\newblock Towards {S}calable {I/O} on a {M}any-core {A}rchitecture.
\newblock In {\em International Conference on Embedded Computer Systems:
  Architectures, MOdeling and Simulation (SAMOS)}, pages 341--348. IEEE, July
  2010.
\newblock \showISBN{978-1-4244-7937-5}.
\newblock \href {http://dx.doi.org/10.1109/ICSAMOS.2010.5642045}
  {\path{doi:10.1109/ICSAMOS.2010.5642045}}.

\bibitem{poss.12}
Raphael `kena' Poss.
\newblock {\em On the realizability of hardware microthreading---Revisting the
  general-purpose processor interface: consequences and challenges}.
\newblock PhD thesis, University of Amsterdam, 2012.
\newblock Available from:
  \url{http://www.raphael.poss.name/on-the-realizability-of-hardware-microthreading/}.

\bibitem{kongetira.05.micro}
P.~Kongetira, K.~Aingaran, and K.~Olukotun.
\newblock Niagara: a 32-way multithreaded {SPARC} processor.
\newblock {\em IEEE Micro}, 25\penalty0 (2):\penalty0 21--29, March/April 2005.
\newblock \showISSN{0272-1732}.
\newblock \href {http://dx.doi.org/10.1109/MM.2005.35}
  {\path{doi:10.1109/MM.2005.35}}.

\bibitem{lankamp.07}
M.~Lankamp.
\newblock {D}eveloping a {R}eference {I}mplementation for a {M}icrogrid of
  {M}icrothreaded {M}icroprocessors.
\newblock Master's thesis, University of Amsterdam, Amsterdam, the Netherlands,
  August 2007.
\newblock Available from:
  \url{http://dist.svp-home.org/doc/mike-lankamp-ref-microgrid.pdf}.

\bibitem{lankamp.12}
Mike Lankamp.
\newblock {\em Design and Evaluation of a Multithreaded Many-Core
  Architecture}.
\newblock PhD thesis, University of Amsterdam, 201x.
\newblock To appear.

\bibitem{mcmahon.86}
F.H. McMahon.
\newblock The livermore {FORTRAN} kernels: A computer test of the numerical
  performance range.
\newblock Technical Report UCRL-53745, Lawrence Livermore National Lab., {CA}
  (USA), Dec 1986.

\bibitem{poss.12.rapido}
Raphael Poss, Mike Lankamp, M.~Irfan Uddin, Jaroslav S\'{y}kora, and Leo\v{s}
  Kafka.
\newblock Heterogeneous integration to simplify many-core architecture
  simulations.
\newblock In {\em Proc. 2012 Workshop on Rapid Simulation and Performance
  Evaluation: Methods and Tools}, RAPIDO '12, pages 17--24. ACM, 2012.
\newblock \showISBN{978-1-4503-1114-4}.
\newblock Available from: \url{pub/poss.12.rapido.pdf}, \href
  {http://dx.doi.org/10.1145/2162131.2162134}
  {\path{doi:10.1145/2162131.2162134}}.

\bibitem{poss.12.dsd}
Raphael Poss, Mike Lankamp, Qiang Yang, Jian Fu, Michiel~W. {van Tol}, and
  Chris Jesshope.
\newblock {Apple-CORE}: {Microgrids} of {SVP} cores (invited paper).
\newblock In Smail Niar, editor, {\em Proc. 15th Euromicro Conference on
  Digital System Design (DSD 2012)}. IEEE Computer Society, September 2012.
\newblock \showISBN{978-0-7695-4798-5}.
\newblock Available from: \url{pub/poss.12.dsd.pdf}, \href
  {http://dx.doi.org/10.1109/DSD.2012.25} {\path{doi:10.1109/DSD.2012.25}}.

\bibitem{zhang.07}
Li~Zhang and Chris~R. Jesshope.
\newblock On-{C}hip {COMA} {C}ache-{C}oherence {P}rotocol for {M}icrogrids of
  {M}icrothreaded {C}ores.
\newblock In Bouge and et~al., editors, {\em Euro-Par Workshops}, volume 4854
  of {\em LNCS}, pages 38--48. Springer, 2007.

\end{thebibliography}

\end{document}